\documentclass[twoside,english]{jfm}
\usepackage[T1]{fontenc}
\usepackage[latin9]{inputenc}
\usepackage[active]{srcltx}
\usepackage{verbatim}
\usepackage{float}
\usepackage{textcomp}
\usepackage{amsmath}
\usepackage{amssymb}
\usepackage{wasysym}
\usepackage{graphicx}
\usepackage{esint}

\makeatletter
\numberwithin{equation}{section}
\numberwithin{figure}{section}

\usepackage{graphicx}
\usepackage{natbib}

\ifCUPmtlplainloaded \else
  \checkfont{eurm10}
  \iffontfound
    \IfFileExists{upmath.sty}
      {\typeout{^^JFound AMS Euler Roman fonts on the system,
                   using the 'upmath' package.^^J}%
       \usepackage{upmath}}
      {\typeout{^^JFound AMS Euler Roman fonts on the system, but you
                   dont seem to have the}%
       \typeout{'upmath' package installed. JFM.cls can take advantage
                 of these fonts,^^Jif you use 'upmath' package.^^J}%
      }
  \else
  \fi
\fi

\ifCUPmtlplainloaded \else
  \checkfont{msam10}
  \iffontfound
    \IfFileExists{amssymb.sty}
      {\typeout{^^JFound AMS Symbol fonts on the system, using the
                'amssymb' package.^^J}%
       \usepackage{amssymb}%
         
         \let\geq=\geqslant
      }{}
  \fi
\fi

\ifCUPmtlplainloaded \else
  \IfFileExists{amsbsy.sty}
    {\typeout{^^JFound the 'amsbsy' package on the system, using it.^^J}%
     \usepackage{amsbsy}}
    {\providecommand\boldsymbol[1]{\mbox{\boldmath $##1$}}}
\fi

\newsavebox{\astrutbox}
\sbox{\astrutbox}{\rule[-5pt]{0pt}{20pt}}

\title[Symmetry reduction for pipe flows]{Symmetry reduction of turbulent pipe flows}

\author[F. Fedele, O. Abessi and P. J. Roberts]%
{Francesco Fedele$^{1,2}$%
  \thanks{Email address for correspondence: fedele@gatech.edu},\ns
Ozeair Abessi$^3$ and Philip J. Roberts$^1$}

\affiliation{$^1$School of Civil and Environmental Engineering, Georgia Institute of Technology,
Atlanta, GA 30322, USA\\[\affilskip]
$^2$School of Electrical and Computer Engineering, Georgia Institute of Technology, Atlanta, GA 30322, USA\\[\affilskip]
$^3$Babol Noshirvani University of Technology, Babol, Iran}

\pubyear{2010}
\volume{650}
\pagerange{119--126}
\date{?; revised ?; accepted ?. - To be entered by editorial office}

\makeatother

\usepackage{babel}
\begin{document}
\maketitle\global\long\def\S{\mathcal{S}}
\global\long\def\eps{\varepsilon}
\global\long\def\H{\mathcal{H}}
\global\long\def\L{\mathcal{L}}
\global\long\def\M{\mathcal{M}}
\global\long\def\K{\mathbf{K}}
\global\long\def\Hilb{\mathbf{H}}
\global\long\def\R{\mathbb{R}}
\global\long\def\ud{\mathrm{d}}

\begin{abstract}
We propose and apply a Fourier-based symmetry reduction scheme to
remove, or quotient, the streamwise translation symmetry of Laser-Induced-Fluorescence
measurements of turbulent pipe flows that are viewed as dynamical
systems in a high-dimensional state space. We also explain the relation
between Taylor's hypothesis and the comoving frame velocity $U_{d}$
of the turbulent orbit in state space. In particular, in physical
space we observe flow structures that deform as they advect downstream
at a speed that differs significantly from $U_{d}$. Indeed, the symmetry
reduction analysis of planar dye concentration fields at Reynolds
number $\mathfrak{\mathsf{Re}}=3200$ reveals that the speed $u$
at which high concentration peaks advect is roughly 1.43 times $U_{d}$.
In a physically meaningful symmetry-reduced frame, the excess speed
$u-U_{d}\approx0.43U_{d}$ can be explained in terms of the so-called
geometric phase velocity $U_{g}$ associated with the orbit in state
space. The ``self-propulsion velocity'' $U_{g}$ is induced by the
shape-changing dynamics of passive scalar structures observed in the
symmetry-reduced frame, in analogy with that of a swimmer at low Reynolds
numbers. \end{abstract}

\section{Introduction}

In the last decade, incompressible fluid turbulence in channel flows
has been studied as chaotic dynamics in the state space of a high-dimensional
system at moderate Reynolds numbers (see, for example, \cite{GibsonetalJFMCouette2008,WillisCvitanovic2013}).
Here, turbulence is viewed as an effective random walk in state space
through a repertoire of invariant solutions of the Navier-Stokes equations
(\cite{CvitanovicJFM2013_clockwork} and references therein). In state
space, turbulent trajectories or orbits visit the neighborhoods of
equilibria, traveling waves or periodic orbits, switching from one
saddle to the other through their stable and unstable manifolds (\cite{CvitanovicPOT_1991},
see also \cite{ChaosBook}). Recent studies on the geometry of the
state space of Kolmogorov flows (\cite{Chandler_Kerswell2013_Kolm})
and barotropic atmospheric models (\cite{Gritsun2011,Gritsun2013})
give evidence that unstable periodic orbits provide the skeleton that
underpins the chaotic dynamics of fluid turbulence. 

In pipe flows, the intrinsic continuous streamwise translation symmetry
and azimuthal symmetry make it difficult to identify invariant flow
structures, such as traveling waves or relative equilibria (\cite{FaisstEckhardt,WedinKerswell2004})
and relative periodic orbits (\cite{VISWANATH2007}), embedded in
turbulence. These structures travel downstream with their own mean
velocity and there is no unique comoving frame that can simultaneously
reduce all relative periodic orbits to periodic orbits and all traveling
waves to equilibria. Recently, this issue has been addressed by \cite{WillisCvitanovic2013}
using the method of slices (\cite{Siminos2011,Froehlich}, see also
\cite{RowleyMarsden2000,RowleyMarsden2003}) to quotient group symmetries
that reveal the geometry of the state space of pipe flows at moderate
Reynolds numbers. Further, \cite{BudanurPRL} exploits the 'first
Fourier mode slice' to reduce the $SO(2)$-symmetry in spatially extended
systems. In particular, they separate the dynamics of the Kuramoto-Shivasinsky
equation into shape-changing dynamics within a quotient or symmetry-reduced
space (base manifold) and a one-dimensional (1-D) transverse space
(fiber) associated with the group symmetry. This is the geometric
structure of a fibration of the state space into a base manifold and
transversal fibers attached to it. Thus, the state space is geometrically
a principal fiber bundle (e.g. \cite{Husemoller},\cite{Steenrod},\cite{Hopf1931}):
a base or quotient manifold of the true dynamics that is not associated
with a drift and has attached transverse fibers of invariant directions. 

In this work, we propose a symmetry reduction for dynamical systems
with translation symmetries, and apply it to symmetry-reduce the evolution
of passive scalars of turbulent pipe flows. The paper is organized
as follows. We first discuss the method of comoving frames for pipe
flows, also referred to as the method of connections (e.g. \cite{RowleyMarsden2000}).
In particular, we explain the relation of comoving frame velocities
to Taylor's (1938) hypothesis. This is followed by an experimental
validation by means of two-dimensional (2-D) Laser-Induced-Fluorescence
(LIF) measurements of planar dye concentration fields of turbulent
pipe flows. The Fourier-based symmetry reduction scheme is then presented
and applied to analyze the acquired experimental data.

\section{Comoving frame velocities and Taylor's hypothesis}

Consider an incompressible three-dimensional (3-D) flow field $\mathbf{\mathbf{v_{\mathrm{0}}}}(x,y,z,t)=(U_{0},V_{0},W_{0})$,
where $x$ and $z$ are the horizontal streamwise and spanwise directions,
and $y$ the vertical axis. The flow satisfies the Navier-Stokes equations
with proper no-slip boundary conditions on generic wall boundaries.
Consider a 3-D passive scalar field $C_{0}(x,y,z,t)$ advected and
dispersed by $\mathbf{v_{\mathrm{0}}}$ in accord with

\begin{equation}
\partial_{t}C_{0}+\mathbf{v_{\mathrm{0}}}\cdot\nabla C_{0}=D_{m}\nabla^{2}C_{0}+f_{0},\label{1}
\end{equation}
where $D_{m}$ is the diffusion coefficient, and $f$ accounts for
sources and sinks. For the pair $(\mathbf{\mathbf{v_{\mathrm{0}}}},C_{0})$,
$\mathbf{\mathbf{v_{\mathrm{0}}}}$ evolves according to the Navier-Stokes
equations with no-slip at the wall boundaries and $C_{0}$ evolves
according to Eq. (\ref{1}). Assume that solutions to both equations
have streamwise translation symmetry. This means that if $\mathbf{(\mathbf{v_{\mathrm{0}}}\mathrm{,\mathit{C}_{0})}}(x,y,t)$
is a solution so is $\mathbf{(\mathbf{v_{\mathrm{0}}}\mathrm{,\mathit{C}_{0})}}(x-\ell,y,t)$
for an arbitrary but fixed shift $\ell$. Hereafter, the translationally
invariant Navier-Stokes velocity field $\mathbf{\mathbf{v_{\mathrm{0}}}}$
is not required to be known or given since our approach is based on
concentration measurements or observables only. 

The presence of translation symmetry allows constructing a symmetry-reduced
system, which (depending on construction) is equivalent to observing
the original system in a comoving frame $(x-\ell_{d}(t),y,z,t)$,
where 
\begin{equation}
U_{d}^{(\mathrm{3D})}=\frac{\mathrm{d}\ell_{d}}{\mathrm{d}t}\label{Ud}
\end{equation}
is the comoving frame velocity for 3-D flows. As a first attempt,
$U_{d}^{(\mathrm{3D})}$ can be chosen to minimize, on average, the
material derivative:

\begin{equation}
\frac{\mathrm{D}C_{0}}{\mathrm{D}t}=\partial_{t}C_{0}+U_{d}^{(\mathrm{3D})}\partial_{x}C_{0},
\end{equation}
namely

\begin{equation}
\left\langle \left(\partial_{t}C_{0}+U_{d}^{(\mathrm{3D})}\partial_{x}C_{0}\right)^{2}\right\rangle _{x,y,z}\label{min}
\end{equation}
is the smallest possible if 
\begin{equation}
U_{d}^{(\mathrm{3D})}(t)=-\frac{\left\langle \partial_{t}C_{0}\partial_{x}C_{0}\right\rangle _{x,y,z}}{\left\langle \left(\partial_{x}C_{0}\right)^{2}\right\rangle _{x,y,z}},\label{Uc}
\end{equation}
where the brackets $\left\langle \cdotp\right\rangle {}_{x,y,z}$
denote space average in $x$, $y$ and $z$. In the comoving frame
$\left(x-\ell_{d}(t),y,z,t\right)$, with $\ell_{d}(t)=\intop_{0}^{t}U_{d}^{(\mathrm{3D})}(\tau)d\tau$,
the passive scalar appears to flow calmly, while still slowly drifting
downstream (see, for example, \cite{KreilosJFM2014} for a study of
parallel shear flows). Only when $\frac{\mathrm{D}C_{0}}{\mathrm{D}t}=0$,
i.e. the diffusion, source and sink terms are in balance, then the
flow is steady in the comoving frame (\cite{KrogstadPoF}), for example
traveling waves (\cite{FaisstEckhardt,WedinKerswell2004}). From (\ref{1}),
(\ref{Uc}) can be written as 
\begin{equation}
U_{d}^{(\mathrm{3D})}(t)=\frac{\left\langle U_{0}\left(\partial_{x}C_{0}\right)^{2}+\partial_{x}C_{0}V_{0}\partial_{y}C_{0}+W_{0}\partial_{x}C_{0}\partial_{z}C_{0}-D_{m}\partial_{x}C_{0}\nabla^{2}C_{0}-f_{0}\partial_{x}C_{0}\right\rangle _{x,y,z}}{\left\langle \left(\partial_{x}C_{0}\right)^{2}\right\rangle _{x,y,z}}.\label{UD3d}
\end{equation}
\begin{comment}
where 
\[
\frac{2B_{c}}{D_{m}}=\left\langle \left(\left.\partial_{x}C_{0}\right|_{x=L}\right)^{2}\right\rangle _{y,z}+-\left\langle \left(\left.\partial_{x}C_{0}\right|_{x=0}\right)^{2}\right\rangle _{y,z}+\left\langle \left(\left.\partial_{x}C_{0}\right|_{x=L}\right)^{2}\right\rangle _{y,z}
\]
is the contribution due to concentration gradients at the inlet ($x=0$)
and outlet ($x=L$) boundaries of the domain of interest. 
\end{comment}
Eq. (\ref{UD3d}) reveals that the comoving frame velocity is a weighted
average of the local flow velocities, sources and sinks. For periodic
boundary conditions the contribution of diffusion processes is null.
From (\ref{Uc}), averaging along the $x$ and $z$ directions only
yields the comoving frame vertical velocity profile 
\begin{equation}
U_{d}^{(\mathrm{3D})}(y,t)=-\frac{\left\langle \partial_{t}C_{0}\partial_{x}C_{0}\right\rangle _{x,z}}{\left\langle \left(\partial_{x}C_{0}\right)^{2}\right\rangle _{x,z}}.\label{Ucpr}
\end{equation}
The associated speed $\widehat{U}_{d}$ of a Fourier mode $\widehat{C_{0}}(k_{x},k_{z,},y,t)e^{i(k_{x}x+k_{z}z)}$
then follows as

\begin{equation}
\widehat{U}_{d}(k_{x},k_{z},y,t)=\frac{\mathrm{Re}\left[i\partial_{t}\widehat{C}_{0}(k_{x},k_{z},y,t)\overline{\widehat{C}}_{0}(k_{x},k_{z},y,t)\right]}{k_{x}\left|\widehat{C_{0}}(k_{x},k_{z},y,t)\right|^{2}},\label{Uck}
\end{equation}
where $\overline{\widehat{C_{0}}}$ is the complex conjugate of $\widehat{C_{0}}$,
$k_{x}$ and $k_{z}$ are the streamwise and crosswise wavenumbers
and $\mathrm{Re}(a)$ denotes the real part of $a$. Note that $\widehat{U}_{d}$
is the same as the convective velocity formulated by \cite{Alamo_JImenez2009}
in the context of Taylor\textquoteright s (1938) \nocite{Taylor1938}abstraction
of turbulent flows as fields of frozen eddies advected by the flow.
When turbulent fluctuations are small compared to the larger scale
flow, they are advected at a speed very close to the time average,
or mean flow velocity $U_{m}$ at a fixed point. %
\begin{comment}
An assumption that advection contributed by turbulent circulations
themselves is small and that therefore the advection of a field of
turbulence past a fixed point can be taken to be entirely due to the
mean flow; also known as the Taylor \textquotedbl{}frozen turbulence\textquotedbl{}
hypothesis.

The time average of the velocity of a fluid at a fixed point, over
a somewhat arbitrary time interval T counted from some fixed time
t0.
\end{comment}
And their temporal variation at frequency $\omega$ at a fixed point
in space can be viewed as the result of an unchanging spatial pattern
of wavelength $2\pi/k_{x}$ convecting uniformly past the point at
velocity $U_{m}=\omega/k_{x}$. This is Taylor's hypothesis that relates
the spatial and temporal characteristics of turbulence. However, eddies
can deform and decay as they are advected downstream and their speed
may differ significantly from $U_{d}^{(\mathrm{3D})}$ and $U_{m}.$

In this regard, \cite{Alamo_JImenez2009} concluded that the comoving
frame or convective velocity $U_{d}^{(\mathrm{3D})}$ of the largest-scale
motion is close to the mean flow speed $U_{m}$, whereas it drops
significantly for smaller-scale motions (\cite{KrogstadPoF}). Hence,
$U_{d}^{(3D)}$ depends on the state of evolution of the flow. For
example, it is well known that turbulent motion in channel flows is
organized in connected regions of the near wall flow that decelerate
and then erupt away from the wall as ejections. These decelerated
motions are followed by larger scale connected motions toward the
wall from above as sweeps. \cite{KrogstadPoF} found that the convection
velocity for ejections is distinctly lower than that for sweeps. 

To gain more insights into the physical meaning of comoving frame
velocities, we performed experiments to trace turbulent pipe flow
patterns using non-intrusive LIF techniques (\cite{TianRoberts2003})
and discussed in the next section.

\subsection{LIF measurements }

\begin{figure}[H]
\centering \includegraphics[width=0.99\textwidth]{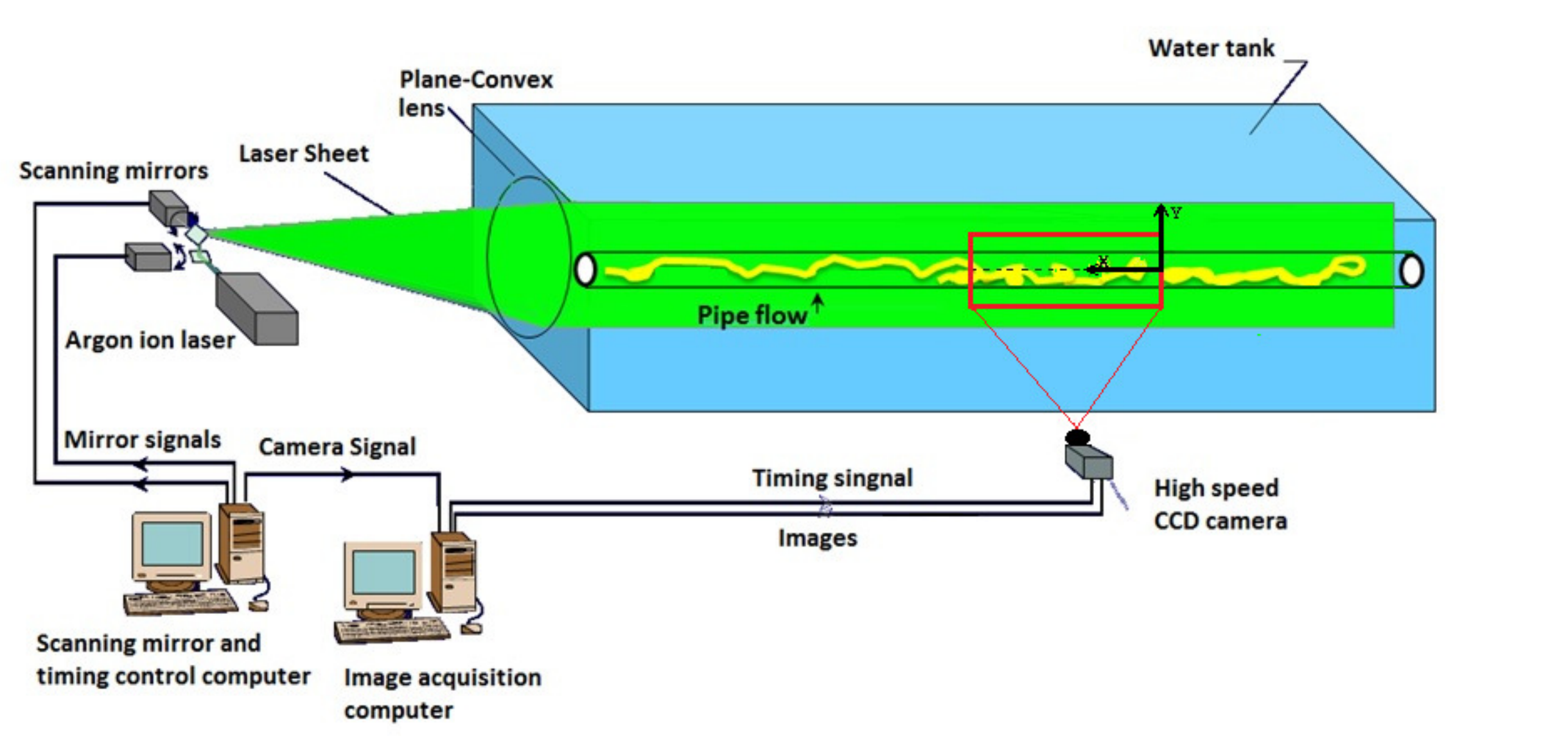}
\protect\caption{Schematic of the LIF system of the Georgia Tech Environmental Fluid
Mechanics Laboratory (\cite{TianRoberts2003}). }

\label{FIGURE1} 
\end{figure}

The experiments were performed in the Environmental Fluid Mechanics
Laboratory at the Georgia Institute of Technology. The LIF configuration
is illustrated in Fig. (\ref{FIGURE1}) and a detailed description
of the system is given in \cite{TianRoberts2003}. The tank has glass
walls $6.10$ m long \texttimes{} $0.91$ m wide \texttimes{} $0.61$
m deep. The front wall consists of two three-meter long glass panels
to enable long unobstructed views. The $5.5$ meter long pipe was
located on the tank floor, and the tank was filled with filtered and
dechlorinated water. The pipe was transparent Lucite tube with radius
$R=2.5\mbox{ cm}$. 

The pipe was completely submerged in water to avoid refraction and
scattering of the emitted light that would occur at the water-Lucite-air
interface along the pipe walls and downstream at the very end of the
pipe when the water flows out with curvy streamlines. With this configuration,
we enable unique LIF imaging of the flow structures in a round pipe
at high flow rate since the pipe discharge is in ambient water instead
of air. 

\begin{figure}[H]
\centering \includegraphics[width=1\textwidth]{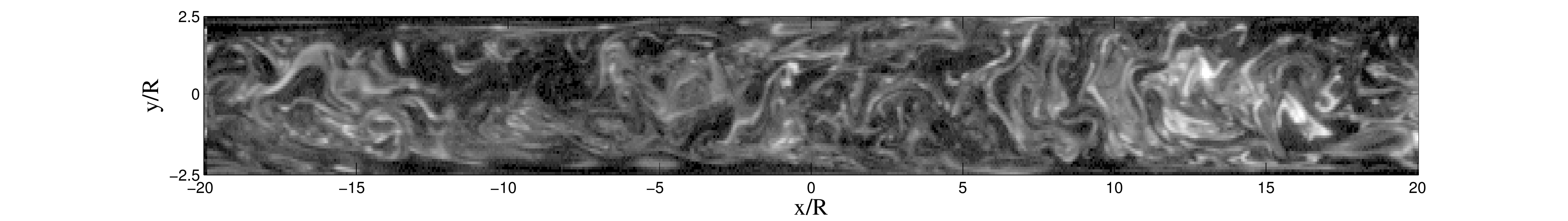} \protect\caption{LIF experiments: snapshot of the planar fluorescent dye concentration
field $C_{0}(x,y,z=0,t)$ tracing turbulent pipe flow patterns at
Reynolds number $\mathsf{Re}=3200$ (bulk velocity $U_{b}=6.42\mbox{ cm/s}$,
flow from right to left). }

\label{FIGURE1a} 
\end{figure}

The water was pumped into a damping chamber to calm the flow, and
then, after passing a rigid polyester filter, it flowed into the pipe.
Fluorescent dye solution was continuously injected into the flow through
a small hole in the pipe wall upstream of the image capture zone of
length $20R$. The solution, a mixture of water and fluorescent dye,
is supplied from a reservoir by a rotary pump at a flowrate measured
by a precision rotameter. The flow was begun and, after waiting a
few minutes for the flow to establish, laser scanning started to record
the experiment. To acquire high resolution data, we captured vertical
centerline planar fluorescent dye concentration fields $C_{0}(x,y,z=0,t)$
which trace turbulent pipe flow patterns. The pipe Reynolds number
$\mathsf{Re}=2U_{b}R/\nu=3200$, where the bulk velocity (discharge
divided by the pipe cross sectional area) $U_{b}=6.42\mbox{ cm/s}$
and $\nu$ is the kinematic viscosity of water. As shown in Fig. (\ref{FIGURE1}),
the vertical laser sheet passes through the pipe centerline to focus
on flow properties in the central plane ($z=0$). Images of the capture
zone ($2R\times20R=5\times50\:\mbox{c\ensuremath{m^{2}}}$) were acquired
at $50$ Hz for $240$ seconds (see Fig. (\ref{FIGURE1a})). The vertical
and horizontal image sizes are $65\times622$ pixels for a resolution
of $0.0794$ cm/pixel.

\begin{figure}[H]
\centering \includegraphics[width=1\textwidth]{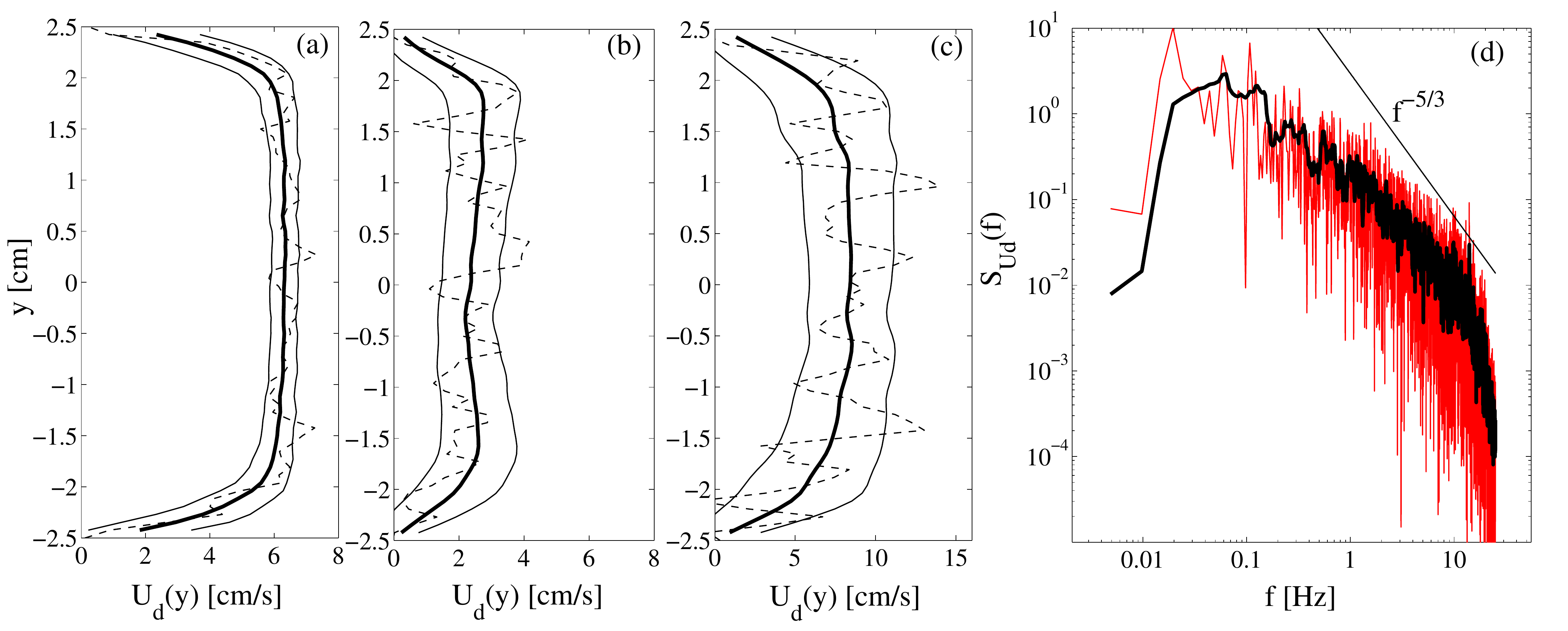}
\protect\caption{LIF experiments: estimated comoving frame, or convective velocity
$U_{d}^{(\mathrm{2D})}(y,t)$ using Eq. (\ref{UDy2D}): time average
profile (solid line), instantaneous profile (dashed line) and its
standard deviations about the mean (thin solid line) estimated accounting
for (a) all spatial scales of the measured $C$ (max speed$=6.32$
cm/s), (b) small scales (max speed$=2.76$ cm/s), and (c) large scales
(max speed$=8.52$ cm/s); (d) observed noisy (red line) and filtered
(black line) frequency spectra of the large-scale comoving frame velocity
$U_{d}^{(\mathrm{2D})}$ {[}see Eq. (\ref{Uc}){]}. Pipe radius $R=2.5$
cm.}

\label{FIGURE2} 
\end{figure}

\begin{figure}[H]
\centering \includegraphics[width=0.8\textwidth]{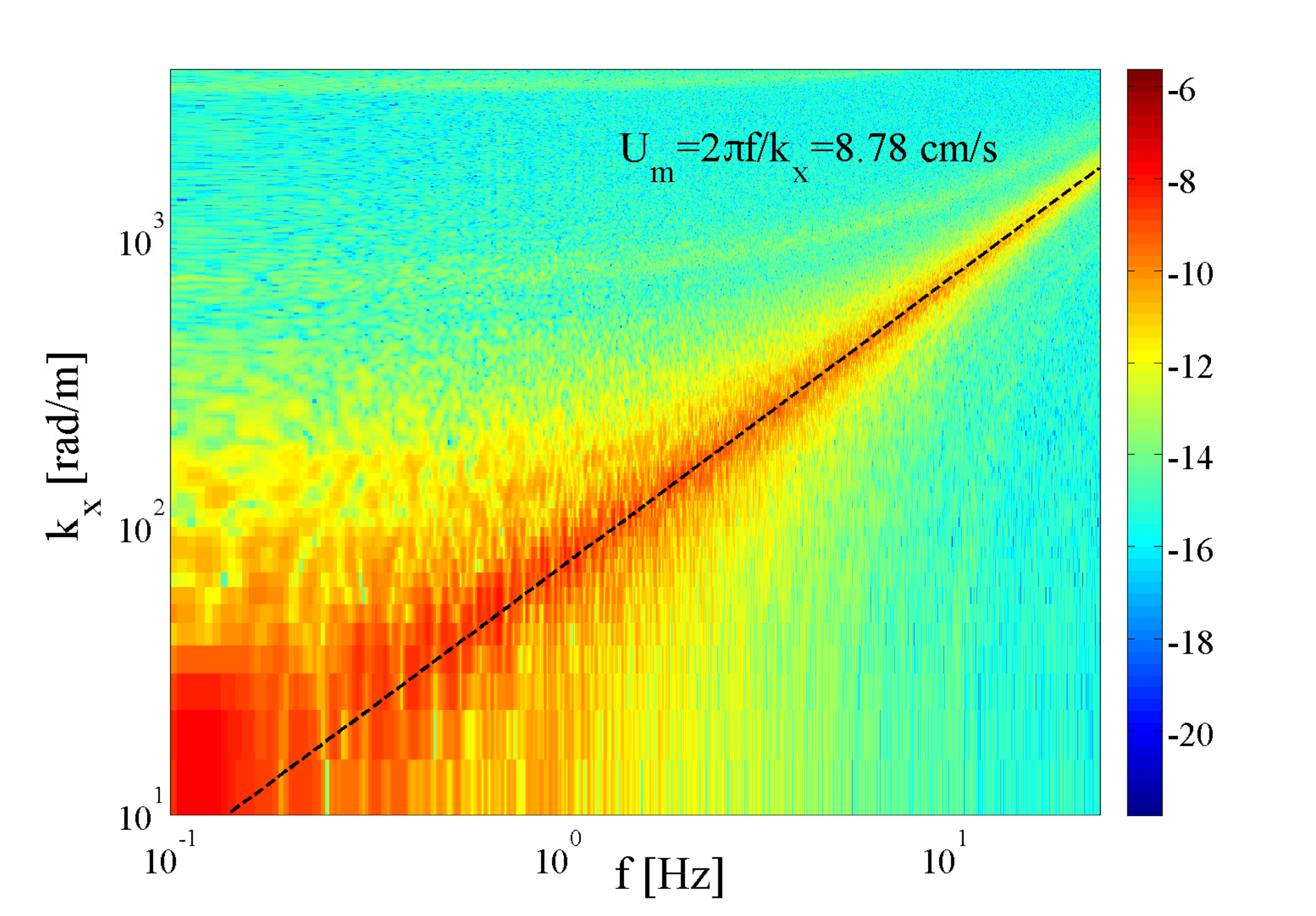}
\protect\caption{Observed log-values of the frequency-wavenumber spectrum $S(k_{x},f)$
of the fluorescent dye concentration $C(x,y=0,t)$ at the pipe centerline.
Estimated mean flow velocity $U_{m}=\omega/k_{x}=2\pi f/k_{x}\sim8.78$
cm/s (dashed line). $U_{m}/U_{b}=1.37$ and bulk velocity $U_{b}=6.42$
cm/s.}

\label{FIGURE3} 
\end{figure}

\begin{figure}[H]
\centering \includegraphics[width=0.99\textwidth]{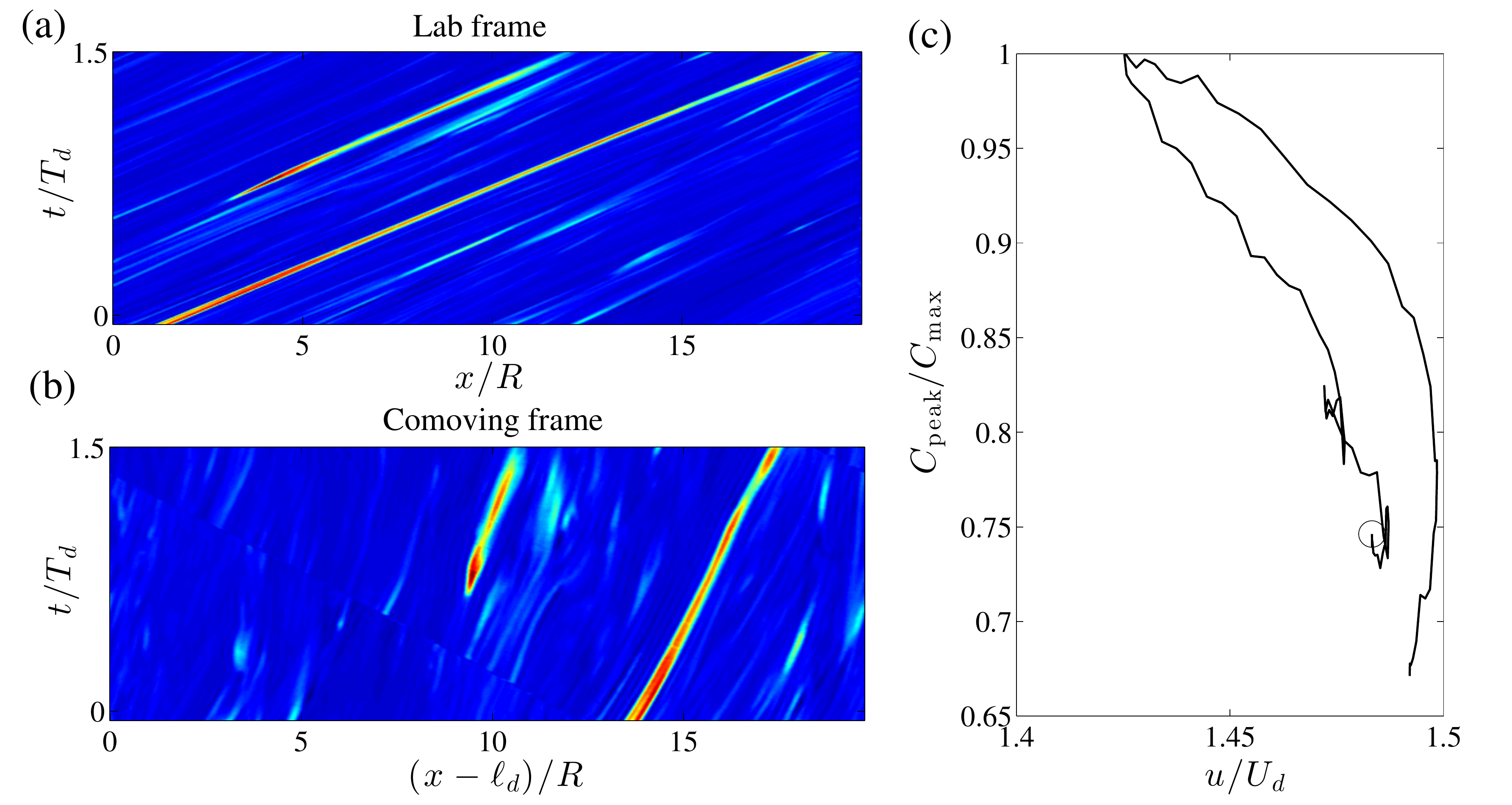}
\protect\caption{LIF experiments: space-time evolution of the dye concentration $C(x,y=0,t)$
at the pipe centerline in the (panel a) lab frame ($x,t)$ and (panel
b) comoving frame $\mbox{\ensuremath{(x-x_{d}(t),t)}}$; (panel c)
normalized instantaneous concentration peak intensity $C_{\mathrm{peak}}/C_{\mathrm{max}}$
tracked from the initial time $t/T_{d}=0$ ($\Circle$) as a function
of the observed peak speed $u/U_{d}^{(2D)}$, with $C_{\mathrm{max}}$
denoting the observed maximum value of $C$ over the whole data set.
$U_{d}\approx$6.34 m/s and $T_{d}=U_{d}/R$.}

\label{FIGURE4} 
\end{figure}

\subsection{Data analysis}

The LIF measurements are planar dye concentration fields $C(x,y,t)=C_{0}(x,y,z=0,t)$
in a vertical slice through the pipe centerline. According to Eq.
(\ref{1}), at $z=0$, the field $C$ satisfies 

\begin{equation}
\partial_{t}C+\mathbf{v_{\mathrm{2D}}}\cdot\nabla_{xy}C=D_{m}\nabla_{xy}^{2}C+f,\label{eq2d}
\end{equation}
where $\nabla_{xy}=\left(\partial_{x},\partial_{y}\right)$ and $\mathbf{v}_{\mathrm{2D}}=(U,V)=(U_{0}(x,y,z=0,t),V_{0}(x,y,z=0,t))$
are the in-plane gradient and flow within the 2D slice and the source
\[
f(x,y,t)=-W_{0}(x,y,z=0,t)\partial_{z}C+D_{m}\left.\partial_{zz}C\right|_{z=0}+f_{0}(x,y,z=0,t)
\]
accounts for the out-of-plane transport and diffusion and in-plane
source/sinks. The associated in-plane comoving frame, or convective,
velocity $U_{d}^{(2D)}$ can be estimated from the measured field
$C(x,y,t)$ using Eq. (\ref{Uc}), where the average is performed
only in the $x$ and $y$ directions, that is
\begin{equation}
U_{d}^{(\mathrm{2D})}(t)=-\frac{\left\langle \partial_{t}C\partial_{x}C\right\rangle _{x,y}}{\left\langle \left(\partial_{x}C\right)^{2}\right\rangle _{x,y}}=\frac{\left\langle \widetilde{U}\left(\partial_{x}C\right)^{2}+\widetilde{V}\partial_{x}C\partial_{y}C-D_{m}\partial_{x}C\nabla_{xy}^{2}C-f\partial_{x}C\right\rangle _{x,y}}{\left\langle \left(\partial_{x}C\right)^{2}\right\rangle _{x,y}}.\label{UD2D}
\end{equation}
Clearly, this depends on the in-plane flow and out-of-plane sources/sinks.
Similarly, the in-plane comoving frame, or convective, velocity profile
$U_{d}^{(\mathrm{2D})}(y,t)$ follows from Eq. (\ref{Ucpr}) averaging
only in the $x$ direction, 
\begin{equation}
U_{d}^{(\mathrm{2D})}(y,t)=-\frac{\left\langle \partial_{t}C\partial_{x}C\right\rangle _{x}}{\left\langle \left(\partial_{x}C\right)^{2}\right\rangle _{x}}.\label{UDy2D}
\end{equation}
For example, Fig. (\ref{FIGURE2}) shows the comoving frame velocity
profiles computed from Eq. (\ref{UDy2D}) including (Panel $a$) all
spatial scales of the measured $C$, (Panel \textbf{$\mathit{b}$})
the small scales (wavelengths $L_{x}<0.2R$, $L_{y}<0.2R$) and (Panel
$c$) the large scales ($L_{x}>2R$, $L_{y}>0.4R$). Clearly, the
small scales advect more slowly than the large scales, in agreement
with \cite{KrogstadPoF}. Moreover, the maximum comoving frame velocity
of the large scales ($=8.52\mbox{ cm/s}$) is close to the centerline
mean flow speed (=$8.78$ cm/s) estimated from the frequency-wavenumber
spectrum of $C(x,y=0,t)$ {[}see Figure (\ref{FIGURE3}){]}. Further,
the frequency spectrum of the comoving frame velocity $U_{d}(t)$
estimated from Eq. (\ref{UD2D}) accounting for large scales only
is also shown in Panel $d$ of Fig. (\ref{FIGURE2}). It decays approximately
as $f^{-5/3}$, indicating that Taylor's hypothesis is approximately
valid, possibly due to the non-dispersive behavior of large scale
motions.

In the fixed frame $(x,t)$, the space-time evolution of the measured
dye concentration $C(x,y=0,t)$ on the pipe centerline is shown in
Panel $a$ of Figure (\ref{FIGURE4}). The associated evolution in
the comoving frame $\left(x-\ell_{d}(t),t\right)$ is shown in Panel
$b$ of the same Figure. The shift $\ell_{d}$ is computed by numerically
integrating $U_{d}^{(\mathrm{2D})}$ in time, which is estimated from
Eq. (\ref{UDy2D}) accounting for all spatial scales of $C$. Note
the shape-changing dynamics of the passive scalar structures, which
still experience a drift in the comoving frame. Moreover, a slowdown
or deceleration is observed in the dye concentration peaks, possibly
related to the above mentioned turbulent flow ejections. This is clearly
seen in Panel $c$ of Figure (\ref{FIGURE4}), which depicts the normalized
instantaneous peak concentration $C_{\mathrm{peak}}$ (normalized
to $C_{\mathrm{max}}$) as a function of the associated peak speed
$u$ (normalized to $U_{d}^{(\mathrm{2D})}$), with $C_{\mathrm{max}}$
denoting the maximum value of $C$ over the whole 2-D data set. Further,
the peak speed $u$ is approximately 40\% larger than the comoving
frame velocity, which is roughly constant during the event ($U_{d}^{(\mathrm{2D})}=6.32\pm0.22$
cm/s). Note that in oceanic wave groups, large focusing crests tend
to slow down as they evolve within the group, as a result of the natural
wave dispersion of unsteady wave trains (\cite{Banner_PRL2014,JFMFedele2014,FedeleEPL2014}).
Thus, we argue that the observed slowdown of the passive scalar peaks
may be due to the wave-like dispersive nature of small-scale turbulent
structures. 

Drawing from differential geometry, the observed excess speed $u-U_{d}^{(\mathrm{2D})}$
of concentration peaks is explained in terms of geometric phases.

\section{Geometric phases}

A classical example in which geometric phases arise is the transport
of a vector tangentially on a sphere. The change in the vector direction
is equal to the solid angle of the closed path spanned by the vector
and it can be described by Hannay's angles (\cite{Hannay}). The rate
at which the angle, or geometric phase, changes in time is the geometric
phase velocity. In physics, the rotation of Foucault\textquoteright s
pendulum can be also explained by means of geometric phases. \cite{Pancharatnam}
discovered their effects in polarized light, and later \cite{Berry08031984}
for quantum-mechanical systems. 

Consider another example drawn from classical mechanics. The dynamics
of a spinning body in a dissipationless media admits rotational symmetry
with respect to the axis of rotation. The associated angular, or geometric
phase velocity $\Omega$ follows from conservation of angular momentum
$I\Omega^{2}$, where $I$ is the moment of inertia. Clearly, $\Omega$
can vary in time if the body shape deforms to induce changes in $I$.
Since the body shape and its deformations are usually known, the rotation
speed depends only on how the shape deforms. Indeed, in the frame
rotating at the speed $\Omega$, we only observe the body shape-changing
dynamics and the rotational symmetry is ``removed'' or quotiented
out. We label this special frame as symmetry-reduced since in a fixed
laboratory frame we cannot distinguish between the body deformation
and spinning motions.

In fluid mechanics, the motion of a swimmer at low Reynolds numbers
can also be explained in terms of geometric phase velocities (\cite{Shapere1}).
In this case, the comoving frame velocity is null since inertia is
neglected and the swimmer's velocity is uniquely determined by the
geometry of the sequence of its body's shapes, which lead to a net
translation, i.e. the geometric phase. In a fixed laboratory frame
we observe the swimmer drifting as its body's shape varies in time,
but it is hard to distinguish between the two types of motions. In
the symmetry-reduced frame moving with the swimmer we only observe
its body deformations and translation symmetry is quotiented out. 

In wave mechanics, the recently noticed slowdown effect of crests
of oceanic wave groups can be explained in terms of geometric phase
velocities (\cite{FedeleEPL2014,Banner_PRL2014}). 

In the abovementioned cases, the associated governing equations are
linear and the shape deformations are known or assumed a priori. Indeed,
in quantum-mechanical systems their shape depends on the eigenfunctions
of the Schrodinger operator (\cite{Berry08031984}). \cite{Shapere1}
considered the eigenfunctions of the Stokes operator to describe the
swimmer's shape. \cite{FedeleEPL2014} considered the special class
of Gaussian envelopes to study the qualitative dynamics of realistic
ocean wave groups. 

In turbulent pipe flows, fluctuating coherent structures advect downstream
at a speed that depends on both their intrinsic properties such as
inertia, and on the way their ``shape'' varies or deforms in time.
However, we don't know a priori their shape as the Navier-Stokes equations
are nonlinear and one cannot rely on an eigenfunction expansion to
model shapes. Clearly, one can use the eigenfunctions of the linearized
Navier-Stokes operator or define a special flow given by the superposition
of patches of constant vorticity whose boundaries change in time according
to given shape modes. However, these are just approximations or simplifications
of the more complex turbulent flows.

In general, the speed of coherent structures includes not only the
comoving frame velocity, which accounts primarily for their inertia,
but also a geometric component. This can be interpreted as a ``self-propulsion''
velocity induced by the shape-changing deformations of the flow structures
similar to that of a swimmer at low Reynolds numbers (\cite{Shapere1}).

To unveil the ``\textit{shape of turbulence}'' we need to quotient
out the translation symmetry. This can be achieved, for example, by
means of a physically meaningful slice representation of the quotient
space (\cite{BudanurPRL,Cvitanovic2012}). Slicing should provide
a symmetry-reduced frame from which one observes the shape-changing
dynamics of coherent structures without drift. The relative velocity
between the comoving and symmetry-reduced frame is the geometric phase
velocity. 

Clearly, in the previous section we have seen that the comoving frame
velocity of pipe flows has the physical meaning of a convective speed.
The geometric phase velocity, on the other hand, depends on an arbitrary
definition of the symmetry-reduced frame. Different slice representations
yield different symmetry-reduced frames, as we will show later. Finding
a physically meaningful symmetry-reduced frame from which one observes
the shape of turbulence is still an open problem. 

In the following, we first present a symmetry reduction scheme to
quotient translation symmetry using slice representations, and then
we apply it to symmetry-reduce the LIF data of turbulent pipe flows
presented in the previous section.

\section{Symmetry reduction via slicing}

As an application, we focus on the desymmetrization of the average
in-plane concentration field $c(x,t)=\left\langle C(x,y,t)\right\rangle {}_{y}$.
It is convenient to express $c$ by means of the truncated Fourier
series

\begin{equation}
\begin{array}[t]{c}
c(x,t)=c_{0}(t)+\frac{1}{2}\sum_{m=1}^{N}z_{m}(t)\exp\left(imk_{x}x\right)+c.c.=\\
\\
c_{0}(t)+\sum_{m=1}^{N}\left|z_{m}(t)\right|\cos\left(mk_{0}x+\theta_{m}(t)\right),
\end{array}\label{cf}
\end{equation}
where $c_{0}(t)$ is the mean, $z_{m}=\left|z_{m}\right|\exp(i\theta_{m})$
is the complex Fourier amplitude with phase $\theta_{m}$, $k_{0}=2\pi/L_{0}$
is a minimum possible wavenumber for the domain length $L_{0}$ of
interest, and the index $m$ runs from $1$ to $N$. The mean $c_{0}$
is invariant under the group action, but its evolution is coupled
to that of the fluctuating component of $c$. This depends on the
evolution of the vector $\mathbf{z}(t)=\{z_{m}\}=(z_{1},...z_{N})$
of Fourier components of $c$ and those of the translationally invariant
Navier-Stokes velocity field $\mathbf{v}$, denoted by the vector
$\hat{\mathbf{v}}$. The velocity field is not required to be given
or known because the proposed symmetry reduction can be applied to
concentration measurements only. %
\begin{comment}
It is straightforward to show that 
\[
\left\langle \left(\partial_{x}c\right)^{2}\right\rangle _{x}=\sum_{m=1}^{N}m^{2}k_{0}^{2}\left|z_{m}\right|^{2}=\left|T(z)\right|^{2}
\]
and
\[
\mathrm{\mathit{\left\langle \partial_{t}c\partial_{x}c\right\rangle _{x}}}=\mathrm{Re}\sum_{m=1}^{N}imk_{0}\overline{z_{m}}\frac{\mathrm{d}z_{m}}{\mathrm{d}t}=\mathrm{-Re}\left(\overline{T(z)}\frac{\mathrm{d}z}{\mathrm{d}t}\right),
\]
and Eq. (\ref{UD1}) can be written as

\begin{equation}
U_{d}(t)=\frac{\mathrm{Re}\left(\overline{T(z)}\frac{\mathrm{d}z}{\mathrm{d}t}\right)}{\left|T(z)\right|^{2}},\label{UDF}
\end{equation}
where we have introduced the vector $T(z)=\left\{ imk_{0}z_{m}\right\} $,
which has a well defined geometric meaning as discussed later on in
the paper {[}see Eq. (\ref{Tz}){]}. 
\end{comment}

The coupled dynamics of $c_{0}$ and $\mathbf{z}$ can be derived
by averaging the governing equation (\ref{eq2d}) in the $y$ direction,
applying flow boundary conditions and projecting onto Fourier basis.
Without losing generality, we can write 
\begin{equation}
\left\{ \begin{array}{c}
\frac{d\mathbf{z}}{dt}=\mathcal{N_{\mathrm{1}}}(c_{0},\mathbf{z},\mathbf{\hat{\mathbf{v}}})\\
\\
\frac{dc_{0}}{dt}=\mathcal{N_{\mathrm{2}}}(c_{0},\mathbf{z},\mathbf{\hat{\mathbf{v}}})
\end{array}\right.,\label{ceq}
\end{equation}
where $\mathcal{N_{\mathrm{1}}}$ and $\mathcal{N_{\mathrm{2}}}$
are appropriate nonlinear operators of their arguments and both are
invariant under translation symmetry, viz. $\mathcal{N_{\mathrm{k}}}(c_{0},g_{\ell}(\mathbf{z}),\mathbf{\mathrm{\mathit{g_{\ell}}}\hat{\mathbf{v}}})=g_{l}\mathcal{N_{\mathrm{k}}}(c_{0},\mathbf{z},\mathbf{\hat{\mathbf{v}}})$.
The orbit $\mathbf{z}$ wanders in the state space $\mathcal{P}\in\mathbb{C^{\mathrm{\mathit{N}}}}$,
and the one-parameter group orbit $g_{\ell}(\mathbf{z})$ of $\mathbf{z}$
is the subspace 

\begin{equation}
g_{\ell}(\mathbf{z})=\left\{ \mathbf{w}\in\mathbb{C^{\mathit{N}}}:\mathbf{w}=\{z_{m}\exp(imk_{0}\ell)\},\:\forall\ell\in\mathbb{R}\right\} ,
\end{equation}
where the length $\ell$ is the drift. For a non-vanishing Fourier
mode $z_{j}$, the symmetry-reduced or desymmetrized orbit $\mathbf{Z}(t)$
is defined by the complex components

\begin{equation}
\mathbf{Z}=\Pi_{j}(\mathbf{z})=\{Z_{m}\}=\left\{ z_{m}\left(\frac{\bar{z}_{j}}{\left|z_{j}\right|}\right)^{m/j}\right\} =\left\{ \left|z_{m}\right|\exp(i\phi_{m})\right\} ,\label{ZD}
\end{equation}
where the phases 
\begin{equation}
\phi_{m}=\theta_{m}-\frac{m\theta_{j}}{j}.\label{phases}
\end{equation}
Note that $Z_{j}=\left|z_{j}\right|$ is real and $\mathbf{Z}\in\mathbb{C^{\mathrm{\mathit{\mathit{N\mathrm{-1}}}}}}$.
For $j=1$, the reduction scheme yields the 'first Fourier mode slice'
proposed in \cite{BudanurPRL}. The scalar field $c_{D}$ in the symmetry-reduced
frame follows from (\ref{cf}) as

\begin{equation}
c_{D}(x,t)=c_{0}(t)+\sum_{m=1}^{N}\left|z_{m}\right|\cos\left(mk_{0}x+\phi_{m}\right).\label{CD}
\end{equation}
It is straightforward to check that any translated copy of $c(x+\ell,t)$
corresponds to a unique $c_{D}$. Indeed, the associated Fourier phases
$\phi_{m}$ in (\ref{phases}) are invariant under the change $\theta_{m}\rightarrow\theta_{m}+m\ell.$
In mathematical terms, the map $\Pi_{j}$ projects an element $\mathbf{z}=\{z_{m}\}$
of $\mathcal{P}$ and all the elements of its group orbit $g_{\ell}\mathbf{\mathrm{(}z\mathrm{)}}$
into the same point $\mathbf{Z}=\Pi_{j}(\mathbf{z})$ of the quotient
space $\mathcal{M}=\mathcal{P}/g_{\ell}\in\mathbb{C^{\mathrm{\mathit{N-1}}}}$,
i.e. $\Pi_{j}(\mathbf{z})=\Pi_{j}(g_{\ell}(\mathbf{z}))$. Note that
$\mathcal{M}$ has one dimension less than the original space since
we have 'removed' translation symmetry. Indeed, $\mathcal{M}$ is
defined as a manifold of $\mathbb{C}^{\mathrm{\mathit{N}}}$ that
satisfies $\mathrm{Im}(z_{j})=0$. 

For $j>1$, the presence of complex roots of $z_{m}$ requires care
in computing the components $Z_{m}$ in (\ref{ZD}). In particular,
we define a slice as a subregion of the original state space $\mathcal{P}$
whose elements are mapped onto the quotient space $\mathcal{M}$ via
the projection map $\Pi_{j}$. Slicing a state space is in general
not unique. In this work, we consider the Fourier slice $S_{j}$ of
$\mathcal{P}$ defined as

\begin{equation}
S_{j}=\left\{ \mathbf{z}\in\mathbb{C^{\mathit{N}}}:z_{j}\neq0\right\} ,\label{Sj}
\end{equation}
which is a region of $\mathbb{C}^{\mathrm{\mathit{N}}}$ delimited
by, but not including, the border of $S_{j}$, i.e. the hyperplane
$z_{j}=0$. $S_{j}$ can be divided in $j$ wedge-shaped subregions
based on the values of the phase $\theta_{j}$ of $z_{j}$ as 
\begin{equation}
S_{j}=\bigcup_{k=0}^{j-1}S_{j,k},
\end{equation}
where the subslice $S_{j,k}$ is the wedge domain defined as 
\begin{equation}
S_{j,k}=\left\{ \mathbf{z}\in\mathbb{C^{\mathit{N}}}:z_{j}\neq0\,\,\,\mathrm{and}\,\,\,\frac{2\pi}{j}k<\theta_{j}<\frac{2\pi}{j}(k+1)\right\} .\label{eq:Sjk}
\end{equation}
The division into subslices is necessary because the phases $\phi_{m}$
of $Z_{m}$ in Eq. (\ref{phases}) jump by $2\pi/j$ each time the
orbit $z_{j}$ winds around the origin of the complex plane crossing
the branch cut $\left\{ \mathrm{Re}(z_{j})\in(0,-\infty)\,\right\} $.
Thus, $\Pi_{j}$ maps elements of $\mathcal{P}$ into any of the $k$
subslices $S_{j,k}$. As $z_{j}$ winds around the origin, a different
$S_{j,k}$ has to be chosen to have continuity of the phases $\phi_{m}$
of $Z_{m}$. Tracking the winding number $\mathrm{Im}\ointop\frac{\mathrm{d}z_{j}}{z_{j}}$
signals when one must switch to a different subslice. In a more practical
way, a jump-free symmetry-reduced orbit $\mathbf{Z}$ is obtained
by first unwrapping the phase $\theta_{j}$ of $z_{j}$ and then computing
$Z_{m}$ by means of Eqs. (\ref{ZD}) and (\ref{phases}). As a result,
$\Pi_{j}$ is defined on the slice $S_{j}$ (see Eq. (\ref{Sj})).

Within the quotient space $\mathcal{M}$, after $j$ cycles are completed
relative equilibria reduce to equilibria and relative periodic orbits
(RPOs) reduce to periodic orbits (POs). Indeed, after one cycle the
projected orbit drifts by $2\pi/(jk_{0})$ in physical space, and
we refer to it as modulo-$2\pi/j$ periodic orbit (MPO). Each RPO
and its shifted copies are uniquely mapped to an MPO in the quotient
space since the symmetry reduction is well defined. Clearly, an ergodic
trajectory, which temporarily visits neighborhoods of RPOs in full
space may experience on average no drift in the desymmetrized or quotient
space if the slice $j$ is properly chosen, as will be shown later
on. The practical and easy choice would be the first Fourier slice
$S_{1}$. However, a good reduction requires the amplitude of $z_{j}$
to be dominant in comparison to the other Fourier components. Indeed,
in general as $z_{j}$ lingers near zero, the orbit wanders near the
border of the slice $S_{j}$. As a result, the map $\Pi_{j}$ becomes
singular since the phase $\theta_{j}$ is undefined (see, for example,
\cite{BudanurPRL}). A different slice can then be chosen and the
slices' borders can be adjoined via ridges into an atlas that spans
the state space region of interest (\cite{Cvitanovic2012}).

The choice of the Fourier slice $S_{j}$ to quotient out the translation
symmetry is entirely arbitrary. Different slices yield different symmetry-reduced
frames in which the concentration field may appear distorted. As an
example, consider the state space to be an infinitely long vertical
cylinder with its vertical lines fibers of the principal bundle (e.g.
\cite{Husemoller},\cite{Steenrod}). Each fiber can be associated
with a single point in the quotient space. If we slice the cylinder
transversally by a plane, the quotient space is an ellipse, or circle
if the plane is orthogonal to the fibers. Of course, we can also slice
the cylinder with a curved surface and the slice is a warped ellipse.
Clearly, different slices are equivalent since slanted/warped ellipses
and circles can be mapped into each other. 

Thus, what is the best Fourier slice representation of turbulent pipe
flows? We argue that a proper choice of the Fourier slice should provide
a physically meaningful symmetry-reduced frame in which the shape-changing
dynamics of coherent structures is observed without drift. In this
case, the observed drift in the comoving frame is explained by means
of geometric phases (see panel $b$ of Fig. (\ref{FIGURE4})). For
our LIF measurements we need to resort to higher order Fourier slices,
as we will show below.

\subsection{Dynamical and geometric phases }

From (\ref{ZD}), the action of the map $\Pi_{j}$ is to shift the
orbit $\mathbf{\mathit{\mathbf{z}}}(t)=\left\{ z_{m}(t)\right\} $
in $\mathcal{P}$ by an amount 
\begin{equation}
\ell_{s}=-\frac{\theta_{j}}{k_{0}j},\qquad j\geq1,\label{shift}
\end{equation}
and the resulting desymmetrized or sliced orbit 
\[
\mbox{\ensuremath{\mathit{\mathbf{\mathrm{\mathbf{Z}}}}}}(t)=g_{-\ell_{s}}(\mathrm{\mathrm{\mbox{\ensuremath{\mathbf{z}}}}})=\left\{ Z_{m}(t)\right\} 
\]
has Fourier components 
\[
Z_{m}=z_{m}\exp(-imk_{0}\ell_{s}),\qquad m=1,...N.
\]
Note that the desymmetrized orbit $\mathbf{Z}=g_{-\ell_{s}}(\mathbf{z})$
does not satisfy the same dynamical equation (\ref{ceq}) for $\mathbf{z}$,
i.e. $\frac{\mathrm{d}\mathbf{z}}{\mathrm{d}t}=\mathcal{N_{\mathrm{1}}}(\mathbf{z})$.
Indeed, (see appendix) 
\begin{equation}
\frac{\mathrm{d}\mathbf{Z}}{\mathrm{d}t}+\frac{\mathrm{d}\ell_{s}}{\mathrm{d}t}T(\mathbf{Z})-\mathcal{N_{\mathrm{1}}}(\mathbf{Z})=0,\label{eqZ}
\end{equation}
where 
\begin{equation}
T(\mathbf{Z})=(g_{\ell}^{-1}\partial_{\ell}g)(\mathbf{Z})=\{imk_{0}Z_{m}\}\label{Tz}
\end{equation}
is the tangent space to the group orbit at $\mathbf{Z}$ (see, for
example, \cite{ChaosBook}).

It is well known that the total drift $\ell_{s}$ is the sum of dynamical
($\ell_{d}$) and geometric ($\ell_{g}$) phase drifts (\cite{Simon,Bhandari})
\begin{equation}
\ell_{s}=\ell_{d}+\ell_{g},\label{shift2}
\end{equation}
where 
\begin{equation}
\ell_{d}=\int_{0}^{t}U_{d}\mathrm{d}\tau,\qquad\qquad\ell_{g}=\int_{0}^{t}U_{g}\mathrm{d}\tau.\label{xd2}
\end{equation}
Here, we have defined the associated dynamical ($U_{d}$) and geometric
($U_{g}$) phase velocities and the total drift speed follows as 
\begin{equation}
U_{s}=\frac{d\ell_{s}}{dt}=U_{d}+U_{g}.\label{Us}
\end{equation}
The decomposition in dynamical and geometric components of the drift
$\ell_{s}$ and associated velocity follows from the condition of
transversality of the symmetry-reduced trajectory $\mathbf{Z}$ to
the group orbit $g_{\ell_{s}}(\mathbf{Z})$, that is $d\mathbf{Z}/dt$
is transversal to the group orbit tangent $T(\mathbf{\mathbf{Z}})$
(\cite{VISWANATH2007,ChaosBook}). Indeed, multiply both members of
Eq. (\ref{eqZ}) by $\overline{T(\mathbf{Z})}$ as 

\begin{equation}
\overline{T(\mathbf{\mathbf{Z}})}\cdot\frac{\mathrm{d}\mathbf{Z}}{\mathrm{d}t}+\frac{\mathrm{d}\ell_{s}}{\mathrm{d}t}\left|T(\mathbf{Z})\right|^{2}-\overline{T(\mathbf{Z})}\mathcal{\cdot N_{\mathrm{1}}}(\mathbf{Z})=0,\label{TT}
\end{equation}
where 
\begin{equation}
\mathbf{a}\cdot\mathbf{b}=\overline{a}_{p}W_{pq}b_{q}\label{dotpr}
\end{equation}
is a weighted scalar product of two vectors with weights $W_{pq}=\overline{W}_{qp}$.
In this work we will use the standard scalar product and the group
orbit is sliced orthogonally, i.e. $W_{pq}=\delta_{pq}$ where $\delta_{pq}$
is the Kronecker symbol. 

The rate of change of the total drift $\ell_{s}$ is a real number
and it follows from the real part of Eq. (\ref{TT}) as 
\begin{equation}
U_{s}=\frac{\mathrm{d}\ell_{s}}{\mathrm{d}t}=\underset{\mathrm{dynamic}}{\underbrace{\frac{\mathrm{Re}\left(\overline{T(\mathbf{Z})}\mathcal{\cdot N_{\mathrm{1}}}(\mathbf{Z})\right)}{\left|T(\mathbf{Z})\right|^{2}}}+}\underset{\mathrm{geometric}}{\underbrace{\frac{\mathrm{-Re}\left(\overline{T(\mathbf{Z})}\cdot\frac{\mathrm{d}\mathbf{Z}}{\mathrm{d}t}\right)}{\left|T(\mathbf{Z})\right|^{2}}}}.\label{dls}
\end{equation}
Here, 
\begin{equation}
U_{d}=\frac{\mathrm{d}\ell_{d}}{\mathrm{d}t}=\frac{\mathrm{Re}\left(\overline{T(\mathbf{Z})}\mathcal{\cdot N_{\mathrm{1}}}(\mathbf{Z})\right)}{\left|T(\mathbf{Z})\right|^{2}}.\label{Ud-2}
\end{equation}
is the so-called dynamical phase velocity (\cite{Simon,Bhandari}).
Since $\overline{T(\mathbf{Z})}\mathcal{\cdot N}_{\mathrm{1}}(\mathbf{Z})$
and $\left|T(\mathbf{Z})\right|^{2}$ are invariant under translation
symmetry, $\ell_{d}$ can also be determined replacing $\mathbf{Z}$
with the orbit $\mathbf{z}$ in $\mathcal{P}$, which is usually known
or observable in applications. Indeed, from (\ref{Ud-2}) and (\ref{ceq})
\begin{equation}
U_{d}=\frac{\mathrm{Re}\left(\overline{T(\mathbf{z})}\mathcal{\cdot N_{\mathrm{1}}}(\mathbf{z})\right)}{\left|T(\mathbf{z})\right|^{2}}=\frac{\mathrm{Re}\left(\overline{T(\mathbf{z})}\cdot\frac{\mathrm{d}\mathbf{z}}{\mathrm{d}t}\right)}{\left|T(\mathbf{z})\right|^{2}}.\label{Ud3}
\end{equation}
It is straightforward to show that $U_{d}$ depends on the evolution
of the concentration field $c$ in the fixed laboratory frame $\mbox{\ensuremath{(x,t)}}$
associated with the orbit $\mathbf{z}$ in $\mathcal{P}$. Indeed,
since
\begin{equation}
\left\langle \left(\partial_{x}c\right)^{2}\right\rangle _{x}=\sum_{m=1}^{N}m^{2}k_{0}^{2}\left|z_{m}\right|^{2}=\left|T(\mathbf{z})\right|^{2}\label{c1}
\end{equation}
and
\begin{equation}
\mathrm{\mathit{\left\langle \partial_{t}c\partial_{x}c\right\rangle _{x}}}=\mathrm{Re}\sum_{m=1}^{N}imk_{0}\overline{z_{m}}\frac{\mathrm{d}z_{m}}{\mathrm{d}t}=\mathrm{-Re}\left(\overline{T(\mathbf{z})}\cdot\frac{\mathrm{d}\mathbf{z}}{\mathrm{d}t}\right),\label{c2}
\end{equation}
it follows that
\begin{equation}
U_{d}=-\frac{\left\langle \partial_{t}c\partial_{x}c\right\rangle _{x}}{\left\langle \left(\partial_{x}c\right)^{2}\right\rangle _{x}}.\label{UDF-1}
\end{equation}
Thus, the dynamical phase velocity $U_{d}$ is the 1-D comoving frame,
or convective, speed similar to that defined for 2-D and 3-D concentration
fields (see Eqs. (\ref{UD2D}) and (\ref{Uc}), respectively). Clearly,
$U_{d}$ also follows by minimization of the spatial mean of the material
derivative of $c$ as in Eq. (\ref{min}). Further, from (\ref{dls})
we define the geometric phase velocity as
\begin{equation}
U_{g}=\frac{\mathrm{d}\ell_{g}}{\mathrm{d}t}=\frac{\mathrm{Re}\left(\overline{T(\mathbf{Z})}\cdot\frac{\mathrm{d}\mathbf{Z}}{\mathrm{d}t}\right)}{\left|T(\mathbf{Z})\right|^{2}}.\label{xg-1}
\end{equation}
Note that $U_{g}$ and $U_{d}$ in (\ref{UDF-1}) are not the same
since $\frac{\mathrm{d}\mathbf{Z}}{\mathrm{d}t}\neq\frac{\mathrm{d}\mathbf{z}}{\mathrm{d}t}$
(see Eqs. (\ref{ceq}) and (\ref{eqZ})). Further, in contrast to
the dynamical $U_{d}$, the geometric $U_{g}$ cannot be related to
the evolution of the concentration field $c$ in the fixed laboratory
frame $\mbox{\ensuremath{\left(x,t\right)}}$; it depends only on
the shape-changing evolution of the desymmetrized field $c_{D}$ (see
Eq. (\ref{CD})) in the symmetry-reduced frame $\left(x-\ell_{d}-\ell_{g},t\right)$.
Here, we recall that $c_{D}$ is associated with the desymmetrized
orbit $\mbox{\ensuremath{\mathbf{Z}}}$ in the quotient space $\mathcal{M}$,
or base manifold. Indeed, Eq. (\ref{xg-1}) can be written as 
\begin{equation}
U_{g}=-\frac{\left\langle \partial_{t}c_{D}\partial_{x}c_{D}\right\rangle _{x}}{\left\langle \left(\partial_{x}c_{D}\right)^{2}\right\rangle _{x}},\label{UgD}
\end{equation}
where we have used Eqs. (\ref{c1},\ref{c2}) replacing $\mathbf{z}$
with $\mathbf{Z}$. Clearly, the geometric phase velocity depends
on the arbitrary choice of the Fourier slice $S_{j}$. Indeed, different
slices yield different desymmetrized concentration fields $c_{D}$,
as discussed later. Further, different scalar products in Eq. (\ref{dotpr})
could be used to filter out the contribution of large or small flow
scales leading to different slice representations. As mentioned above,
in this work we only consider the standard scalar product and all
flow scales are accounted for. 

The comoving orbit 
\begin{equation}
\mathbf{Z}_{d}(t)=g_{-\ell_{d}}(\mathbf{z})\label{comovingorbit}
\end{equation}
is the orbit seen from a comoving frame drifting at the speed $U_{d}$.
In physical space it corresponds to an evolution of the dye concentration
in the comoving frame $\left(x-\ell_{d},t\right)$. Note that in general
$\ell_{d}(t)$ is time varying, and constant only for traveling waves.
Clearly, the dynamical drift $\ell_{d}$ increases with the time spent
by the trajectory $\mathbf{z}(t)$ to wander around $\mathcal{P}$.
The geometric drift $\ell_{g}$ instead depends upon the path $\Gamma=\{Z_{n}(t)\}$
associated with the desymmetrized orbit $\mathbf{Z}(t)$ in the quotient
space. Indeed, 
\begin{equation}
\ell_{g}(t)=\int_{0}^{t}U_{g}\mathrm{d}\tau=-\int_{0}^{t}\frac{\mathrm{Re}\left(\overline{T(\mathbf{Z})}\cdot\frac{\mathrm{d}\mathbf{Z}}{\mathrm{d}\tau}\right)}{\left|T(\mathbf{Z})\right|^{2}}\mathrm{d}\tau=-\int_{\Gamma}\frac{\mathrm{Re}\left(\overline{T(\mathbf{Z})}\mathrm{\cdot d}\mathbf{Z}\right)}{\left|T(\mathbf{Z})\right|^{2}}.\label{xg}
\end{equation}
The desymmetrized orbit $\mathbf{Z}$ is obtained by further shifting
the comoving orbit $\mathbf{Z}_{d}$ in Eq. (\ref{comovingorbit})
by the geometric drift $\ell_{g}$ as 
\begin{equation}
\mathbf{Z}=g_{-\ell_{g}}(\mathbf{Z}_{d})=g_{-\ell_{d}-\ell_{g}}(\mathbf{z}).\label{Z}
\end{equation}
Different slice representations yield different symmetry-reduced frames
$(x-\ell_{d}-\ell_{g},t)$ from which one observes distorted shape-changing
dynamics of the dye concentration field. Only relative equilibria
or traveling waves have null geometric phase, since their shape is
not dynamically changing in the base manifold as they reduce to equilibria.
The geometric drift $\ell_{g}$ and associated speed $U_{g}$ can
be indirectly computed from Eqs. (\ref{shift2},\ref{Us}) as $\ell_{g}=\ell_{s}-\ell_{d}$
and $U_{g}=U_{s}-U_{d}$ respectively. The pairs $(\ell_{s}$,$U_{s}=d\ell_{s}/dt)$
and $(\ell_{d}=\int_{0}^{t}U_{d}\mathrm{d}\tau,U_{d})$ are easily
estimated from concentration measurements. 

The Fourier slice should be properly chosen to provide a physically
meaningful symmetry-reduced frame, as discussed in the next section.

\begin{figure}[H]
\centering \includegraphics[width=1.15\textwidth]{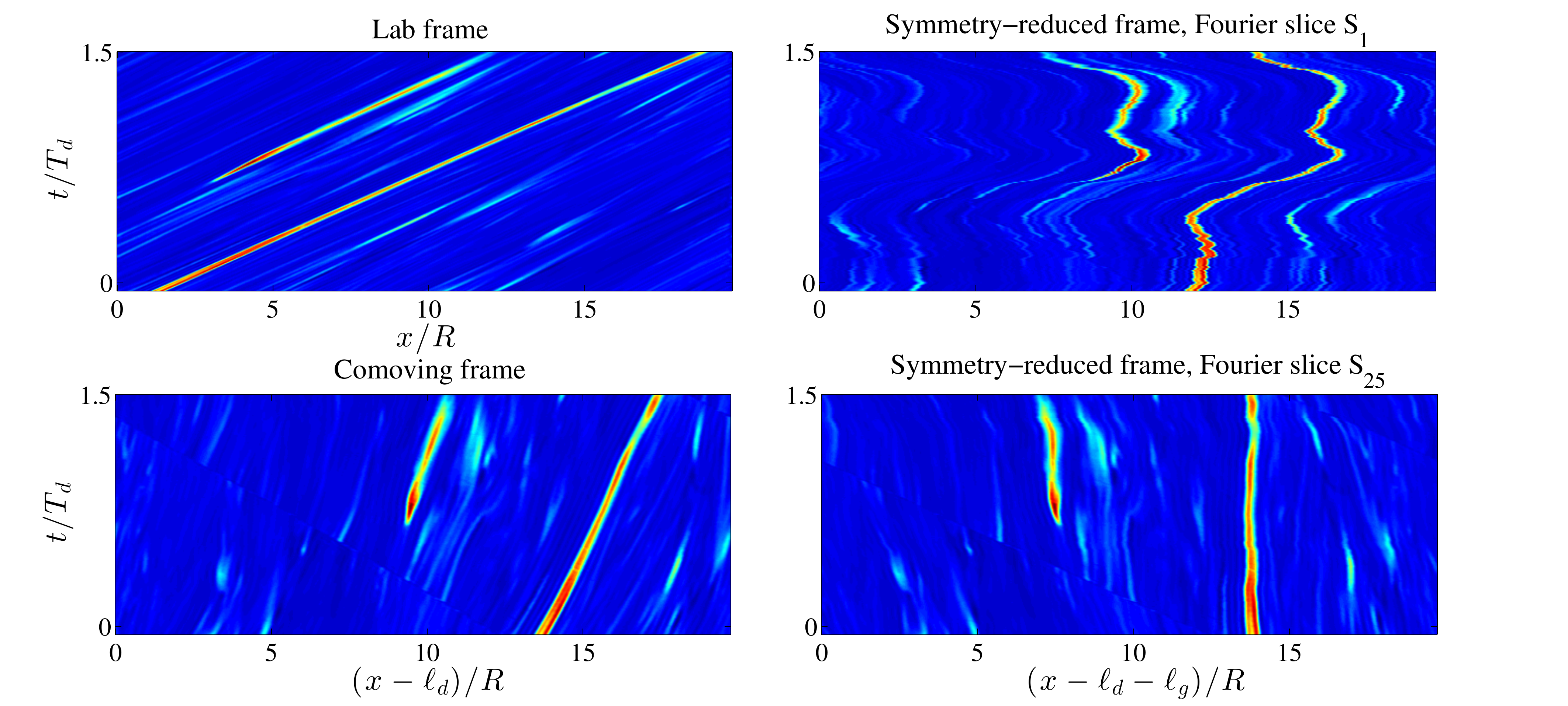}
\protect\caption{Symmetry reduction of LIF measurements: space-time evolution of a
passive scalar structures; (left panels) measured concentration $C(x,y=0,t)$
at the pipe centerline in the (left-top) fixed frame, (left-bottom)
comoving frame $\left(x-\ell_{d},t\right)$, (right panels) symmetry-reduced
frame $\left(x-\ell_{d}-\ell_{g},t\right)$ using Fourier slices (right-top)
$S_{1}$ and (right-bottom) $S_{25}$; time average $\overline{U}_{d}\approx6.74$
cm/s, $\overline{U}_{g}\approx0.4U_{d}$ and $T_{d}=\overline{U}_{d}/R$.}

\label{FIGURE6} 
\end{figure}

\begin{figure}[H]
\centering \includegraphics[width=1.15\textwidth]{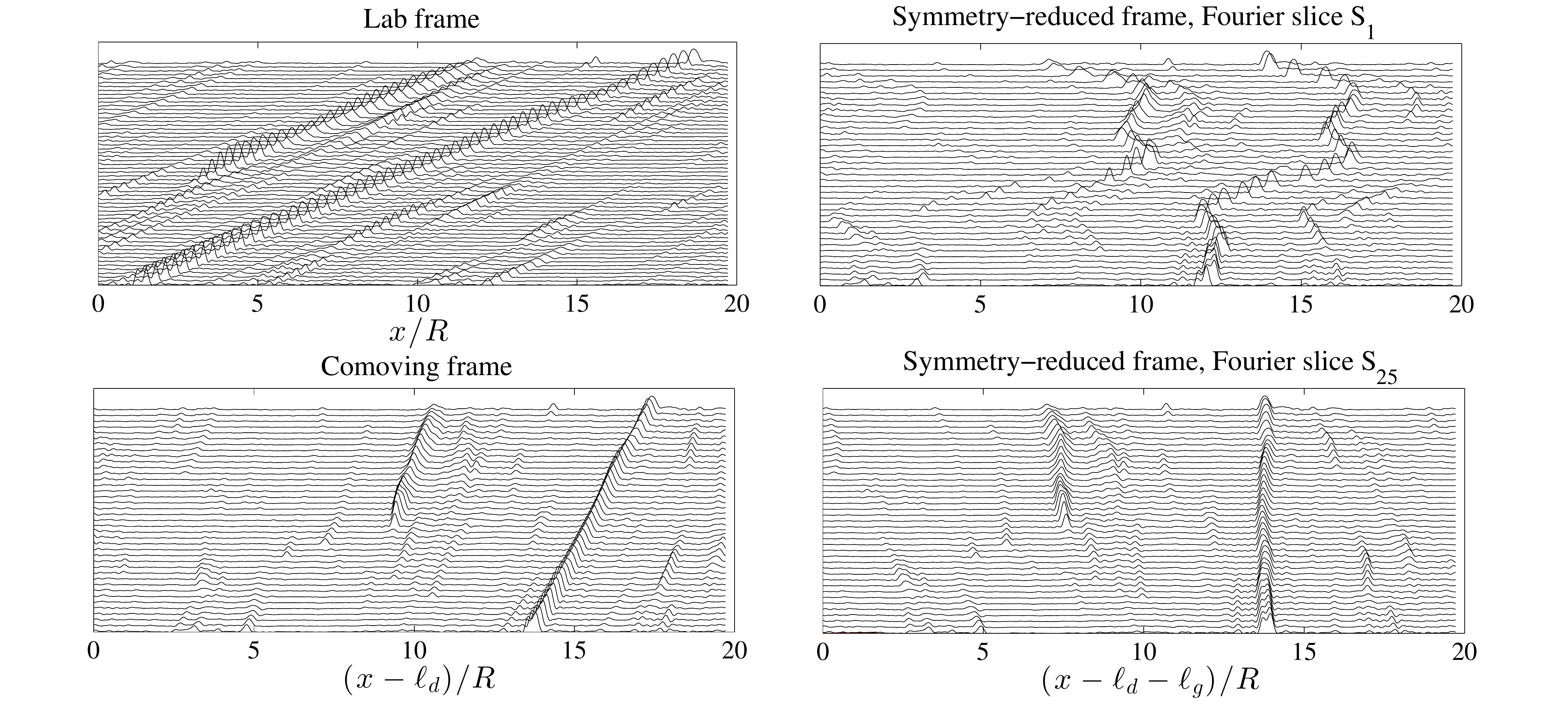}
\protect\caption{Symmetry reduction of LIF measurements: concentration profiles at
increasing instants of time of the measured concentration $c(x,y=0,t)$
at the pipe centerline in the (top-left panel) fixed frame, (bottom-left
panel) comoving frame $\left(x-\ell_{d},t\right)$, (right panels)
symmetry-reduced frame $\left(x-\ell_{d}-\ell_{g},t\right)$ using
Fourier slices (top) $S_{1}$ and (bottom) $S_{25}$. In each plot
time increases from bottom to top. Associated 2-D patterns are shown
in Fig. (\ref{FIGURE6}).}

\label{FIGURE8a} 
\end{figure}

\begin{comment}
\begin{figure}[h]
\centering \includegraphics[width=1\textwidth,bb = 0 0 200 100, draft, type=eps]{patternchanging_singleevent4_zoom.eps}
\protect\caption{Shape-changing evolution of the measured concentration $c(x,y=0,t)$
at the pipe centerline in the symmetry-reduced frame $\left(x-\ell_{s},t\right)$
(time increases from bottom to top).}

\label{FIGURE8} 
\end{figure}
\end{comment}

\begin{figure}[H]
\centering \includegraphics[width=1.12\textwidth]{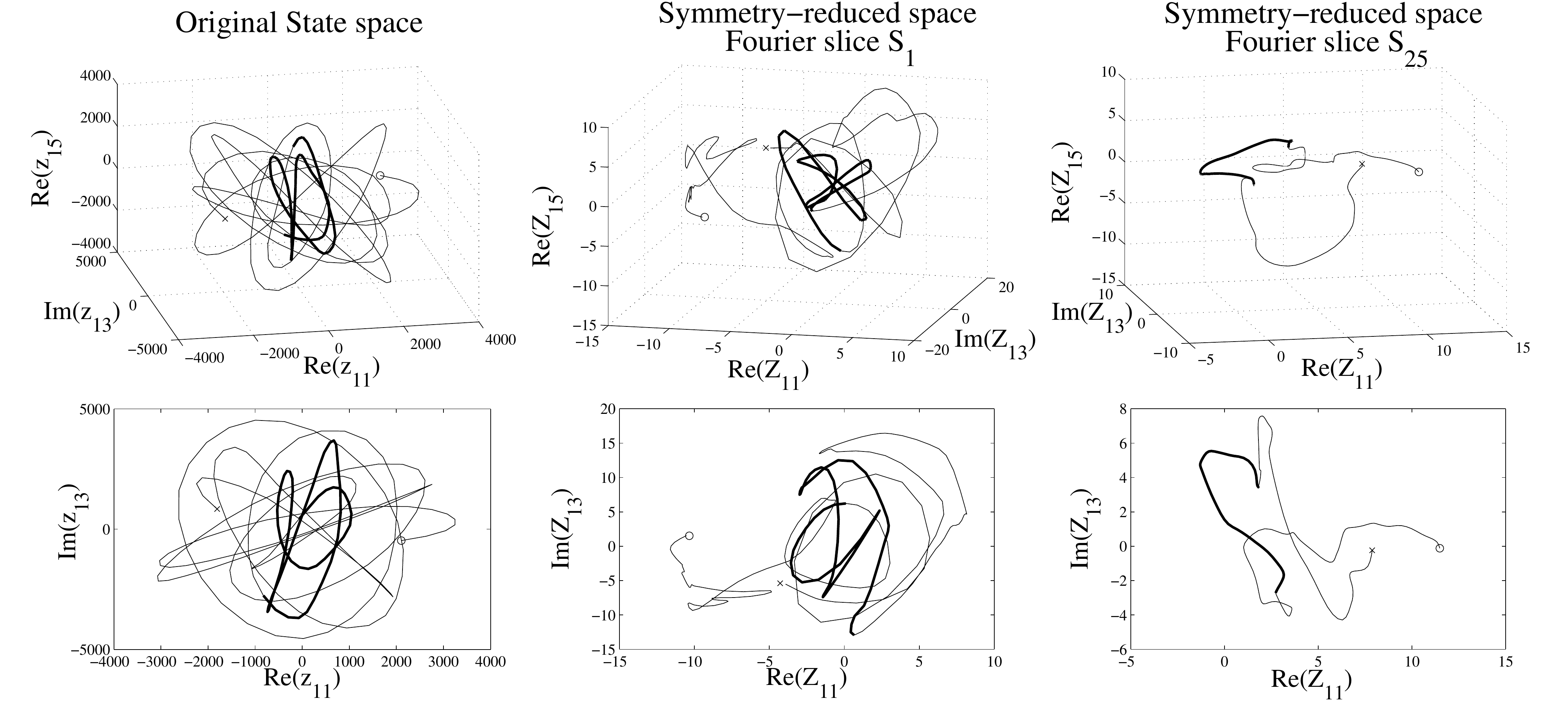}
\protect\caption{Symmetry reduction of LIF measurements: (left panels) Orbit trajectory
$\mathbf{z}$ in the subspace $\left\{ \mbox{Re}(z_{11}),\mbox{Im}(z_{13}),\mbox{Re}(z_{15})\right\} $
of the state space $\mathcal{P}$ associated with the passive scalar
dynamics in the lab frame of Fig. (\ref{FIGURE8a}) (see also Fig.
(\ref{FIGURE6})); (center panels) corresponding symmetry-reduced
orbits $\mathbf{Z}$ in the subspace $\left\{ \mbox{Re}(Z_{11}),\mbox{Im}(Z_{13}),\mbox{Re}(Z_{15})\right\} $
of the base manifold $\mathcal{M}$ associated with Fourier slices
(center panels) $S_{1}$ and (right panels) $S_{25}$. The bold line
indicates the excursion of the orbit while the concentration $c$
lingers above the threshold $0.95c_{\mathrm{max}}$ ($\Circle$ =initial
time, $\times$=final time).}

\label{FIGURE7} 
\end{figure}

\begin{figure}[H]
\centering \includegraphics[width=0.8\textwidth]{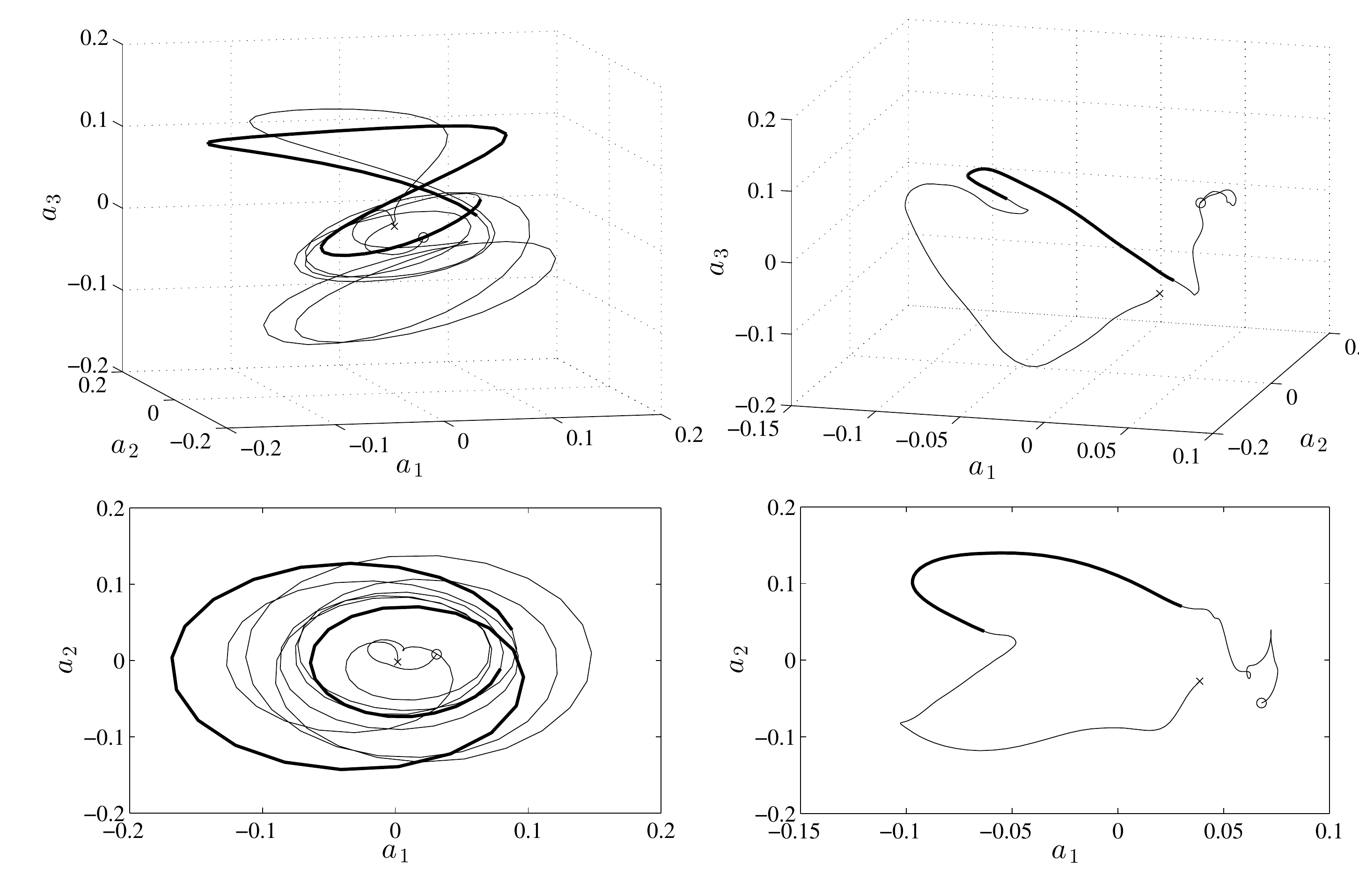}
\protect\caption{Symmetry reduction of LIF measurements: (left) Orbit trajectories
$\mathbf{z}$ associated with the passive scalar dynamics in the lab
frame (see panel (a) of Fig. (\ref{FIGURE6})) projected onto the
subspace ($a_{1},a_{2},a_{3}$) of the most energetic POD modes (right)
corresponding desymmetrized orbit $\mathbf{Z}$ in the symmetry-reduced
frame associated with the Fourier slice $S_{25}$. The bold line indicates
the excursion of the orbit while the concentration $c$ lingers above
the threshold $0.95c_{\mathrm{max}}$ ($\Circle$ =initial time, $\times$=final
time).}

\label{FIGURE12} 
\end{figure}

\subsection{Symmetry reduction of LIF measurements}

In the section, we present a symmetry reduction of the acquired LIF
measurements of turbulent pipe flows (see section 2.1). In particular,
we study their evolution in physical space and in the associated state
space $\mathcal{P}$ of dimension $N=40430$ equal to the total number
of data image pixels ($65\times622$). 

As regards the choice of the Fourier slice $S_{j}$ it is in general
entirely arbitrary. There is no unique way to quotient out the symmetry.
The most likely choice would be $S_{1}$, but for our measurements
this choice will not produce a physically meaningful symmetry reduction.
Higher order slices are required.

In particular, Figs. (\ref{FIGURE6}) and (\ref{FIGURE8a}) illustrate
the space-time evolution of a passive scalar structure and concentration
profiles. The top-left panel of Fig. (\ref{FIGURE6}) shows the dye
concentration $c(x,t)$ at the pipe centerline in the fixed frame
$(x,t)$ {[}see also Fig. (\ref{FIGURE8a}){]}. A drift in the streamwise
direction $x$ is observed. The corresponding orbit $\mathbf{z}(t)$
in the subspace $\left\{ \mbox{Re}(z_{11}),\mbox{Im}(z_{13}),\mbox{Re}(z_{15})\right\} $
of $\mathcal{P}$ is shown in the left panels of Fig. (\ref{FIGURE7}).
Note that the excursion of the orbit while the concentration $c$
lingers above the threshold $0.95c_{\mathrm{max}}$ is complicated
(bold line) since it wanders around its group orbit as a result of
the drift induced by the translation symmetry. The bottom-left panel
of Fig. (\ref{FIGURE6}) shows the space-time evolution in the comoving
frame $\left(x-\ell_{d},t\right)$. Note that the dye concentration
still experiences a significant drift (see also Fig. (\ref{FIGURE8a})).
As a result, the associated orbit in state space still wanders around
the group orbit. A proper choice of the Fourier slice can provide
a physically meaningful symmetry-reduced frame. For example, if we
choose the first Fourier mode slice $S_{1}$, the top-left panel of
Fig. (\ref{FIGURE6}) depicts the associated evolution in the symmetry-reduced
frame $\left(x-\ell_{d}-\ell_{g},t\right)$. Clearly, the symmetry
is quotient out, but in the Fourier slice $S_{1}$ we observe a distorted
shape-changing dynamics of the dye concentration. Instead, if we choose
the Fourier slice $S_{25}$ the drift almost disappears in the corresponding
symmetry-reduced frame, as shown in the bottom-left panel of Figure
(\ref{FIGURE6}) {[}see also Fig. (\ref{FIGURE8a}){]}. Here, this
slice is sufficient to symmetry-reduce the orbit $\mathbf{z}$ over
the analyzed time span as its Fourier components $z_{k},$ with $k\sim20-30$,
never lingers near zero, whereas smaller or larger wavenumber modes
can be small. %
\begin{comment}
The slice $S_{25}$ is the best choice for a physically meaningful
shape-changing dynamics of the passive scalar structures in the symmetry-reduced
frame, as clearly seen in bottom panel of Figure (\ref{FIGURE8a}). 
\end{comment}
The corresponding symmetry-reduced orbits $\mathbf{Z}(t)$ associated
with $S_{1}$ and $S_{25}$ are computed from Eq. (\ref{ZD}). Their
time evolutions within the subspace $\left\{ \mbox{Re}(Z_{11}),\mbox{Im}(Z_{13}),\mbox{Re}(Z_{15})\right\} $
of $\mathcal{M}$ are shown in the center and left panels of Fig.
(\ref{FIGURE7}) respectively. Here, the excursion of the orbits while
the concentration $c$ is high ($>0.95c_{\mathrm{max}}$) is marked
as a bold line. Similar dynamics is also observed projecting the orbits
onto the subspace of their respective most energetic proper orthogonal
decomposition (POD) modes, as shown in Fig. (\ref{FIGURE12}). The
POD projection of the symmetry-reduced orbit $\mathbf{Z}$ is performed
within the corresponding symmetry-reduced space. Note that any two
POD mode amplitudes are statistically uncorrelated by construction
as any two components $Z_{p}$ and $Z_{q}$ chosen at random. Clearly,
this does not imply that they are stochastically independent since
they evolve on the quotient manifold $\mathcal{M}$, which is unknown.
As an example, consider two random variables $X$ and $Y$ that satisfy
$X^{2}+Y^{2}-1=0$. They are uncorrelated but not independent and
POD projections will not help revealing the intrinsic manifold structure.
Local linear embedding techniques may be more appropriate and appealing
(\cite{LLE}), but they are beyond the scope of our work.

\begin{figure}[H]
\centering \includegraphics[width=1\textwidth]{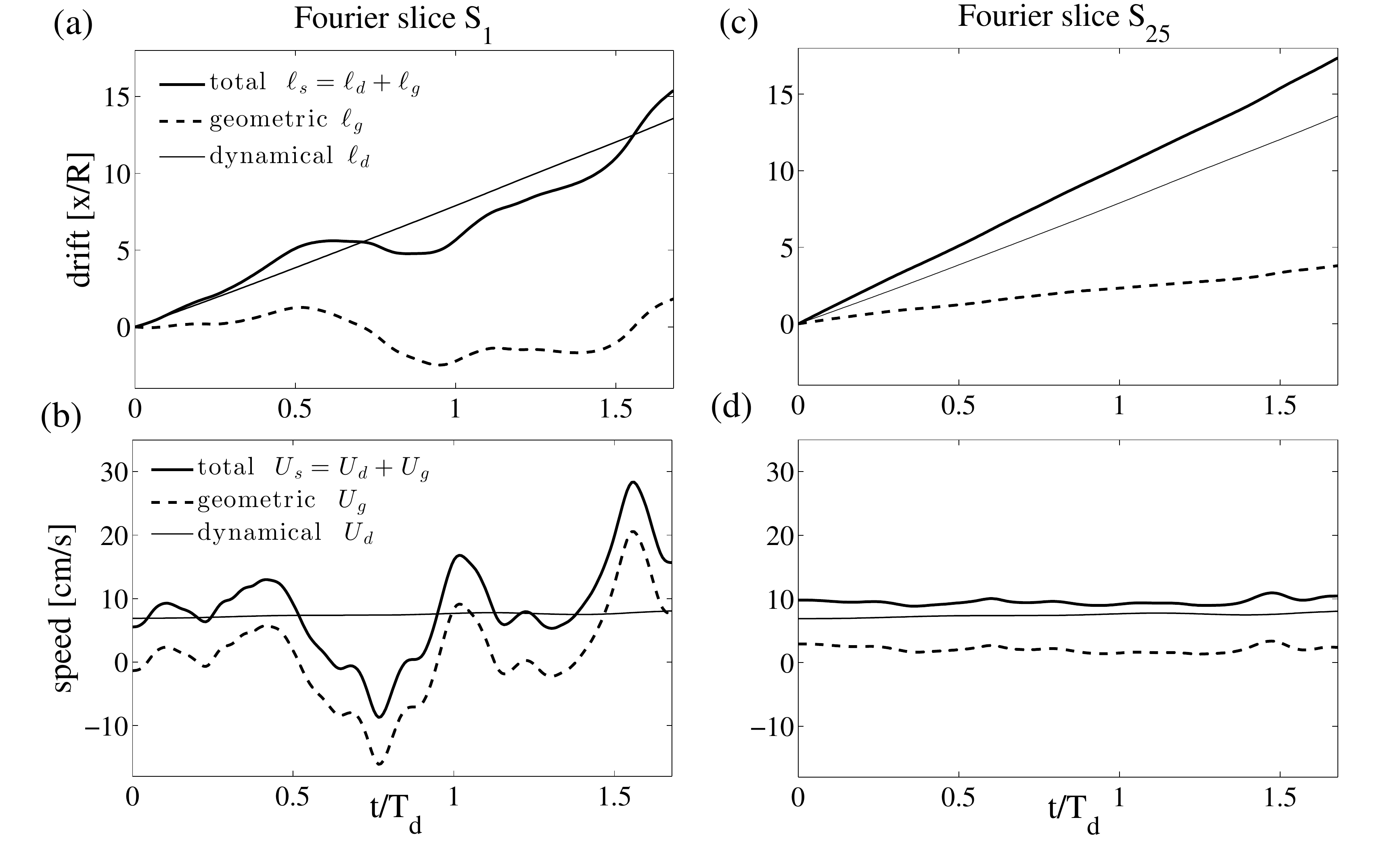}
\protect\caption{Symmetry reduction of LIF data using Fourier slice $S_{1}$ (panels
a-b) and $S_{25}$ (panels $c,d$); (top panels $a,c$) total, dynamical
and geometric drifts and (bottom panels $b,d$) corresponding velocities
$U_{s}$, $U_{d}$ and $U_{g}$ associated with the orbit in state
space of Fig. (\ref{FIGURE7}).}

\label{FIGURE9} 
\end{figure}

The top panels $a,c$ of Figure (\ref{FIGURE9}) show that geometric
drifts associated with Fourier slices $S_{1}$ and $S_{25}$ are different
and so the respective geometric phase velocities (see bottom panels
$b,d$ of the same figure). Clearly, the dynamical component $U_{d}$
is the same since it does not depend on the symmetry-reduction scheme
or slice. Note that $\ell_{g}$ is not the drift seen by an observer
in the symmetry-reduced frame. If it were, the geometric phase velocity
associated with the slice $S_{25}$ would be zero since the desymmetrized
dye concentration field does not drift. If the same observer drifts
by $\ell_{g}$ he will observe the dynamics in the comoving frame.
This explains why the geometric phase velocity $U_{g}$ associated
with the slice $S_{1}$ is negative between the time span $0.5<t/T_{d}<1$.
With reference to Fig. (\ref{FIGURE8a}), in that time interval an
observer in the symmetry-reduced frame needs to decelerate in order
to follow the dye concentration evolution seen in the comoving frame. 

\begin{figure}[h]
\centering \includegraphics[width=1\textwidth]{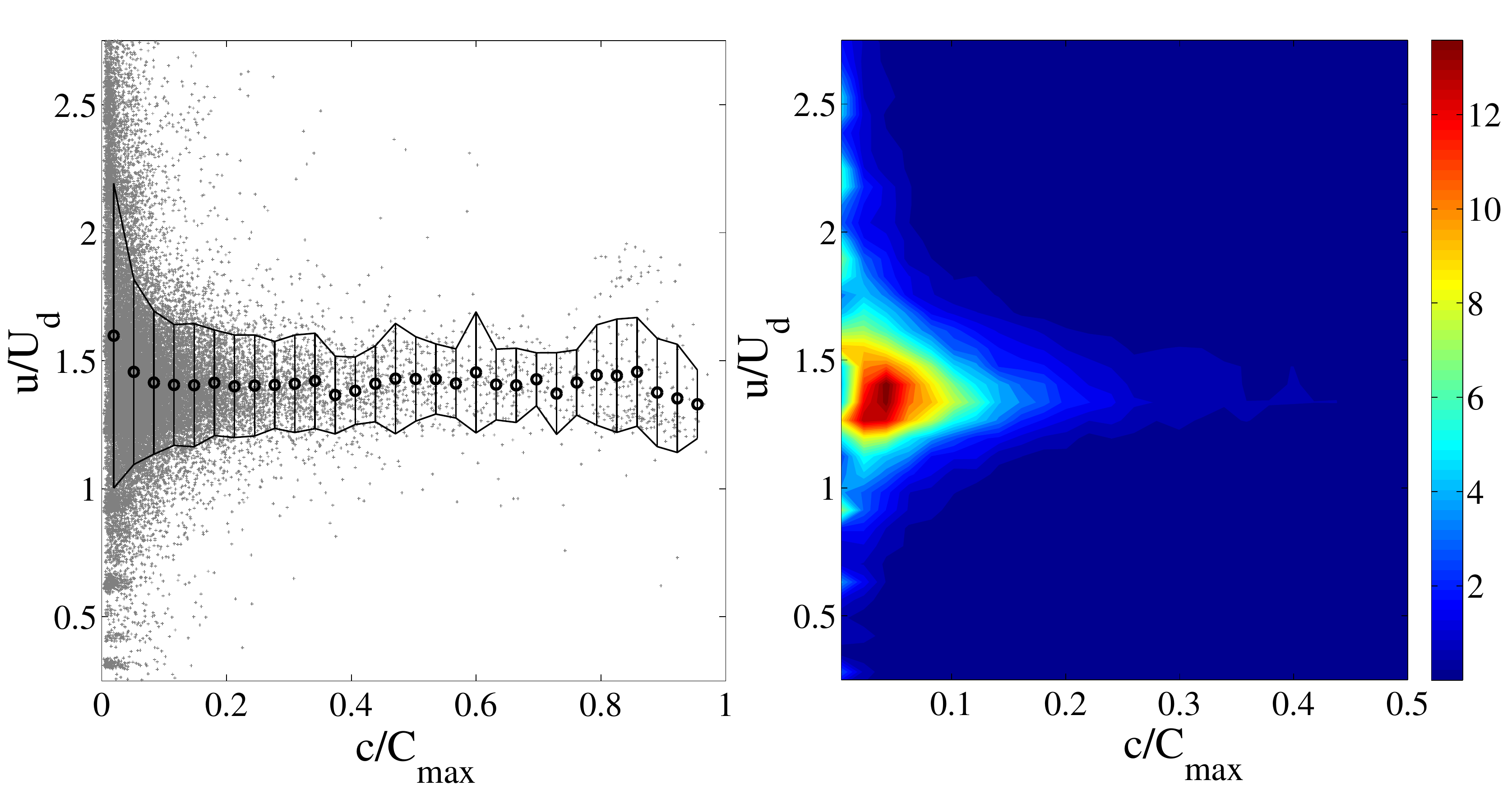}
\protect\caption{LIF experiments: (left) observed normalized dye concentration peak
speed $u/U_{d}$ as a function of the amplitude peak $c/C_{\mathrm{max}}$,
and (right) associated probability density function, with $C_{\mathrm{max}}$
denoting the observed maximum value of dye concentration over the
whole data set. }

\label{FIGURE10} 
\end{figure}

The observed speed $u$ of dye concentration peaks is approximately
40\% larger than the comoving frame velocity $U_{d}$, which changes
slightly during the event. The excess speed $\delta u=u-U_{d}$ is
fairly explained by the geometric phase velocity $U_{g}\approx0.4U_{d}$
associated with slice $S_{25}$, as seen in panel $d$ of Figure (\ref{FIGURE9}).
This appears to be a general trend of the flow as can be seen in Fig.
(\ref{FIGURE10}), which shows the observed normalized speed $u/U_{d}$
of dye concentration peaks tracked in space as a function of their
amplitude $c/C_{\mathrm{max}}$, and the associated probability density
function, where $C_{\mathrm{max}}$ denotes the observed maximum value
of dye concentration over the whole data set. As the peak amplitude
increases, their speed $u$ tends to $1.43U_{d}$. Furthermore, in
the symmetry-reduced frame, we observe the shape-changing dynamics
of passive scalar structures (see bottom panel of Figure (\ref{FIGURE8a})).
This induces the 'self-propulsion velocity' $U_{g}$ of the flow structures
similar to that of the motion of a swimmer at low Reynolds numbers
(\cite{Shapere1}). Only when the geometric $U_{g}\ll U_{d}$, Taylor's
approximation is valid and, as a result the flow structures slightly
deform as they are advected at the comoving frame or dynamical phase
velocity $U_{d}$, which is close to the mean flow $U_{m}$. 

\begin{comment}
This induces a 'self-propulsion velocity' $U_{g}$ of the flow structures
similar to that of the motion of a swimmer at low Reynolds numbers.
In this case, the dynamical phase velocity $U_{d}$ is null since
inertia is neglected and the swimmer's velocity is uniquely determined
by the geometry of the sequence of its body's shapes, which lead to
a net translation. This can be explained in terms of geometric phases
(\cite{Shapere1}). Only when these are small, i.e. $U_{g}\ll U_{d}$,
Taylor's approximation is valid and, as a result the flow structures
slightly deform as they are advected at the comoving frame speed $U_{d}$. 
\end{comment}

\section{Conclusions}

We have presented a Fourier-based symmetry reduction scheme for dynamical
systems with continuous translation symmetries. As an application,
we have symmetry-reduced LIF measurements of fluorescent dye concentration
fields tracing a turbulent pipe flow at Reynolds number $\mathsf{Re}=3200$.
The symmetry reduction of LIF data on higher order Fourier slices
revealed that the motion of passive scalar structures is associated
with the dynamical and geometric phases of the corresponding orbits
in state space. In particular, the observed speed $u\approx1.43U_{d}$
of dye concentration peaks exceeds the comoving or convective velocity
$U_{d}$. A physically meaningful representation of the quotient space
by a proper choice of the Fourier slice explains the excess speed
$\delta u=u-U_{d}$ as the geometric phase velocity $U_{g}\approx0.43U_{d}$
associated with the Fourier slice $S_{25}$. Similar to the motion
of a swimmer at low Reynolds number, the excess speed $\delta u$
is a 'self-propulsion' velocity $U_{g}$ induced by the shape-changing
dynamics of passive scalar structures as revealed in the symmetry-reduced
frame. 

Symmetry reduction is promising for the analysis of three-dimensional
LIF and PIV measurements as well as simulated flows of pipe turbulence,
in order to unveil the ``shape of turbulence'' and the hidden skeleton
of its chaotic dynamics in state space. Further, the dependence of
geometric phase velocities on the Reynolds number may shed some light
on the nature of transition to turbulence, since the geometric phase
is a measure of the curvature of the quotient manifold. %
\begin{comment}
Further studies along these directions are desirable. 
\end{comment}

\section{Acknowledgments}

FF acknowledges the Georgia Tech graduate courses ``Classical Mechanics
II'' taught by Jean Bellissard in Spring 2013 and ``Nonlinear dynamics:
Chaos, and what to do about it?{}`` taught by Predrag Cvitanovi\'c
in Spring 2012. FF also thanks Alfred Shapere for discussions on geometric
phases, and Federico Bonetto, Nazmi Burak Budanur, Bruno Eckhardt,
Mohammad Farazmand, Chongchun Zeng as well as Evangelos Siminos for
discussions on symmetry reduction.

\section{Appendix }

The time derivative of $\mathbf{z}=g_{\ell_{s}}(\mathbf{Z})$ is 
\[
\frac{\mathrm{d}\mathbf{z}}{\mathrm{d}t}=g_{\ell_{s}}\left(\frac{\mathrm{d}\mathbf{Z}}{\mathrm{d}t}\right)+\frac{\mathrm{d}\ell_{s}}{\mathrm{d}t}(\partial_{\ell_{s}}g)\mathbf{Z},
\]
and the governing equation (\ref{ceq}) for $\mathbf{z}$ yields 
\[
g_{\ell_{s}}\left(\frac{\mathrm{d}\mathbf{Z}}{\mathrm{d}t}\right)+\frac{\mathrm{d}\ell_{s}}{\mathrm{d}t}(\partial_{\ell_{s}}g)\mathbf{Z}-\mathcal{N_{\mathrm{1}}}(g_{\ell_{s}}\mathbf{Z})=0,
\]
where the dependence of $\mathcal{N_{\mathrm{1}}}$ on $c_{0}$ and
$\mathbf{\hat{\mathbf{\mathrm{\mathbf{v}}}}}$ is dropped for clarity
of notation. Factoring out $g_{\ell_{s}}$ yields 

\[
g_{\ell_{s}}\left(\underset{T(\mathbf{Z})}{\frac{\mathrm{d}\ell}{\mathrm{d}t}\underbrace{g_{\ell_{s}}^{-1}(\partial_{\ell_{s}}g)\mathbf{Z}}}+\frac{\mathrm{d}\mathbf{Z}}{\mathrm{d}t}-\underset{\mathcal{N_{\mathrm{1}}}(\mathbf{Z})}{\underbrace{g_{\ell_{s}}^{-1}\mathcal{N_{\mathrm{1}}}(g_{\ell_{s}}\mathbf{Z})}}\right)=0.
\]
This can be further simplified using (\ref{Tz}) and noting that $\mathcal{N_{\mathrm{1}}}$
is invariant under translation symmetry, that is 
\[
g_{\ell_{s}}\left(\frac{\mathrm{d}\mathbf{Z}}{\mathrm{d}t}+\frac{\mathrm{d}\ell_{s}}{\mathrm{d}t}T(\mathbf{Z})-\mathcal{N_{\mathrm{1}}}(\mathbf{Z})\right)=0.
\]
For translation symmetries, $g_{\ell_{s}}(q)=0$ if and only if $q=0$,
thus the evolution of $\mathbf{Z}$ is governed by
\[
\frac{\mathrm{d}\mathbf{Z}}{\mathrm{d}t}+\frac{\mathrm{d}\ell_{s}}{\mathrm{d}t}T(\mathbf{Z})-\mathcal{N_{\mathrm{1}}}(\mathbf{Z})=0.
\]

\bibliographystyle{jfm}
\bibliography{geometricphases_citationstest}

\end{document}